\documentclass[camera,letterpaper,nomarginnotes,nonarrowgutter]{jpaper}
\usepackage{mathptmx} % This is Times font

\usepackage{cite}
\usepackage{amsmath,amssymb,amsfonts,bm}
\usepackage{algorithmic}
\usepackage{graphicx}
\usepackage{siunitx}
\usepackage{textcomp}
\usepackage{xcolor}
\usepackage{fancyhdr}
\usepackage[normalem]{ulem}
\usepackage{glossaries} % For acronyms / abbreviations
\usepackage{todonotes} % For margin notes
\usepackage{marginnote} 
\let\marginpar\marginnote
\usepackage{setspace}
\usepackage{siunitx} % To use SI unit system
\usepackage{listings} % For code snippets
\usepackage{dblfloatfix}    % To enable figures at the bottom of page
\usepackage[vlined, linesnumbered]{algorithm2e}
\usepackage[flushleft]{threeparttable}

\usepackage{balance}
\usepackage{titlesec}
\usepackage[bookmarks=true,breaklinks=true,hidelinks]{hyperref}
\usepackage{soul}
\DeclareSIUnit[number-unit-product = ]\percent{\char`\%}

\def\BibTeX{{\rm B\kern-.05em{\sc i\kern-.025em b}\kern-.08em
    T\kern-.1667em\lower.7ex\hbox{E}\kern-.125emX}}

\fancypagestyle{firstpage}{
  \fancyhf{}
  
  \fancyfoot[C]{\thepage}
}

\newcommand{\linebreakand}{%
  \end{@IEEEauthorhalign}
  \hfill\mbox{}\par
  \mbox{}\hfill\begin{@IEEEauthorhalign}
}

\newcommand{\affilETH}{$^1$}
\newcommand{\affilTOBB}{$^2$}

\newcommand{\affilCESGA}{$^3$}
\newcommand{\affilETHCESGA}{$^{1,3}$}

\title{HiRA: Hidden Row Activation\\for Reducing Refresh Latency of Off-the-Shelf DRAM Chips} 
\author{
\fontsize{11}{12}\selectfont
{A. Giray Ya\u{g}l{\i}k\c{c}{\i}\affilETH{}}\qquad%
{Ataberk Olgun\affilETH{}}\qquad
{Minesh Patel\affilETH{}}\qquad
{Haocong Luo\affilETH{}}\qquad
{Hasan Hassan\affilETH{}}\\
\fontsize{11}{12}\selectfont
{Lois Orosa\affilETHCESGA{}}\quad
{O\u{g}uz Ergin\affilTOBB{}}\quad
{Onur Mutlu\affilETH{}}\\
{\fontsize{10}{11}\selectfont
\affilETH\emph{ETH Z{\"u}rich}\qquad
\affilTOBB{}TOBB University of Economics and Technology\qquad
\affilCESGA\emph{Galicia Supercomputing Center~(CESGA)}%
}}

\def\BibTeX{{\rm B\kern-.05em{\sc i\kern-.025em b}\kern-.08em
    T\kern-.1667em\lower.7ex\hbox{E}\kern-.125emX}}

\def\UrlBreaks{\do\/\do-\/\do.\/\do:}

\expandafter\def\expandafter\UrlBreaks\expandafter{\UrlBreaks
  \do\a\do\b\do\c\do\d\do\e\do\f\do\g\do\h\do\i\do\j
  \do\k\do\l\do\m\do\n\do\o\do\p\do\q\do\r\do\s\do\t
  \do\u\do\v\do\w\do\x\do\y\do\z\do\A\do\B\do\C\do\D
  \do\E\do\F\do\G\do\H\do\I\do\J\do\K\do\L\do\M\do\N
  \do\O\do\P\do\Q\do\R\do\S\do\T\do\U\do\V\do\W\do\X
  \do\Y\do\Z}
  
\definecolor{amber}{rgb}{1.0, 0.49, 0.0}
\definecolor{awesome}{rgb}{1.0, 0.13, 0.32}
\definecolor{dollarbill}{rgb}{0.52,0.73,0.4}
\definecolor{moegi}{rgb}{0.357, 0.537, 0.188}
\definecolor{burgundy}{rgb}{0.5, 0.0, 0.13}
\definecolor{ballblue}{rgb}{0.13, 0.67, 0.8}
\definecolor{ups-truck}{rgb}{0.53, 0.28, 0.21}
\definecolor{airforceblue}{rgb}{0.36, 0.54, 0.66}
\definecolor{cadmiumgreen}{rgb}{0.0, 0.42, 0.24}
\definecolor{darkcyan}{rgb}{0.0, 0.55, 0.55}
\definecolor{caribbeangreen}{rgb}{0.0, 0.8, 0.6}
\definecolor{flamingopink}{rgb}{0.99, 0.56, 0.67}
\definecolor{jazzberryjam}{rgb}{0.65, 0.04, 0.37}
\definecolor{mediumpersianblue}{rgb}{0.0, 0.4, 0.65}
\definecolor{coolblack}{rgb}{0.0, 0.18, 0.39}
\definecolor{bleudefrance}{rgb}{0.19, 0.55, 0.91}
\definecolor{ao}{rgb}{0.0, 0.0, 1.0}
\definecolor{babyblueeyes}{rgb}{0.63, 0.79, 0.95}
\definecolor{darkwarmgray}{rgb}{0.2,0,0}
\definecolor{brightpink}{rgb}{1.0, 0.0, 0.5}

\newcommand{\circled}[1]{{\tikz[baseline=(char.base)]{\node[shape=circle,inner sep=1pt,fill=black, text=white] (char) {\footnotesize \textbf{#1}};}}}

\newcommand{\squishlist}{
 \begin{list}{$\circ$}
  { \setlength{\itemsep}{0pt}
     \setlength{\parsep}{0pt}
     \setlength{\topsep}{0pt}
     \setlength{\partopsep}{0pt}
     \setlength{\leftmargin}{1em}
     \setlength{\labelwidth}{1em}
     \setlength{\labelsep}{0.5em} } }

\newcommand{\squishsublist}{
\begin{list}{$\rightarrow$}
 { \setlength{\itemsep}{0pt}
    \setlength{\parsep}{0pt}
    \setlength{\topsep}{-10em}
    \setlength{\partopsep}{-3pt}
    \setlength{\leftmargin}{1em}
    \setlength{\labelwidth}{1em}
    \setlength{\labelsep}{0.5em} } }

\newcommand{\squishend}{
  \end{list}  }

\newcommand{\head}[1]{\noindent\textbf{#1.}} 

\newcounter{obs}
\newcounter{take}
\newif\ifrevision
\revisionfalse

\newif\ifsubmission
 \submissionfalse
\submissiontrue

\newif\ifiscarev
\iscarevfalse

\newif\ifmicrorev
\microrevfalse

\newif\ifshepherdone
\shepherdonefalse

\newcommand{\micrev}[1]{#1}
\ifmicrorev
  \renewcommand{\micrev}[1]{\textcolor{blue}{#1}}
\fi

\newcommand{\sho}[1]{#1}
\newcommand{\shom}[1]{#1}
\ifshepherdone
  \renewcommand{\sho}[1]{\textcolor{blue}{#1}}
  \renewcommand{\shom}[1]{\textcolor{blue}{#1}}
\fi

\newif\ifshepherdtwo
\shepherdtwofalse

\ifshepherdtwo
  
\fi

\newif\ifcrb
\crbfalse

\newcommand{\agycrbtodo}[1]{}

\newcommand{\agycrbcomment}[1]{}
\newcommand{\atacrbcomment}[1]{}
\newcommand{\omcrbcomment}[1]{}
\ifcrb
  \renewcommand{\agycrbtodo}[1]{\todo[size=\footnotesize,color=blue]{\textbf{TODO:} #1}}
  \renewcommand{\agycrbcomment}[1]{\todo[size=\footnotesize,color=blue]{\textbf{@gy:} #1}}
  \renewcommand{\atacrbcomment}[1]{\todo[size=\footnotesize,color=blue]{\textbf{@atab:} #1}}
  \renewcommand{\omcrbcomment}[1]{\todo[size=\footnotesize,color=blue]{\textbf{@gy:} #1}}

\fi

\newif\ifcri
\crifalse

\newcommand{\agycritodo}[1]{}
\newcommand{\agycri}[1]{#1}

\newcommand{\agycricomment}[1]{}
\newcommand{\atacricomment}[1]{}
\newcommand{\omcricomment}[1]{}
\ifcri
  \renewcommand{\agycritodo}[1]{\todo[size=\footnotesize,color=red]{\textbf{TODO:} #1}}
  \renewcommand{\agycricomment}[1]{\todo[size=\footnotesize,color=babyblueeyes]{\textbf{@gy:} #1}}
  \renewcommand{\atacricomment}[1]{\todo[size=\footnotesize,color=babyblueeyes]{\textbf{@atab:} #1}}
  \renewcommand{\omcricomment}[1]{\todo[size=\footnotesize,color=burgundy]{\textbf{@gy:} #1}}
  \renewcommand{\agycri}[1]{\textcolor{blue}{#1}}

\fi

\newif\ifcru
\crufalse

\newcommand{\agycrutodo}[1]{}
\newcommand{\agycru}[1]{#1}

\newcommand{\agycrucomment}[1]{}
\newcommand{\atacrucomment}[1]{}
\newcommand{\omcrucomment}[1]{}
\ifcru
  \renewcommand{\agycrutodo}[1]{\todo[size=\footnotesize,color=red]{\textbf{TODO:} #1}}
  \renewcommand{\agycrucomment}[1]{\todo[size=\footnotesize,color=babyblueeyes]{\textbf{@gy:} #1}}
  \renewcommand{\atacrucomment}[1]{\todo[size=\footnotesize,color=babyblueeyes]{\textbf{@atab:} #1}}
  \renewcommand{\omcrucomment}[1]{\todo[size=\footnotesize,color=burgundy]{\textbf{@gy:} #1}}
  \renewcommand{\agycru}[1]{\textcolor{amber}{#1}}

  \renewcommand{\agycru}[1]{\textcolor{blue}{#1}}
\fi

\newif\ifcrd
\crdfalse

\newcommand{\agycrdtodo}[1]{}

\newcommand{\agycrdcomment}[1]{}
\newcommand{\atacrdcomment}[1]{}
\newcommand{\omcrdcomment}[1]{}
\ifcrd
  \renewcommand{\agycrdtodo}[1]{\todo[size=\footnotesize,color=red]{\textbf{TODO:} #1}}
  \renewcommand{\agycrdcomment}[1]{\todo[size=\footnotesize,color=babyblueeyes]{\textbf{@gy:} #1}}
  \renewcommand{\atacrdcomment}[1]{\todo[size=\footnotesize,color=babyblueeyes]{\textbf{@atab:} #1}}
  \renewcommand{\omcrdcomment}[1]{\todo[size=\footnotesize,color=red]{\textbf{onur:} #1}}

\fi

\newif\ifcra
\crafalse

\newcommand{\agycratodo}[1]{}
\newcommand{\agycra}[1]{#1}

\newcommand{\agycracomment}[1]{}
\newcommand{\atacracomment}[1]{}
\newcommand{\omcracomment}[1]{}
\ifcra
  \renewcommand{\agycratodo}[1]{\todo[size=\footnotesize,color=red]{\textbf{TODO:} #1}}
  \renewcommand{\agycracomment}[1]{\todo[size=\footnotesize,color=babyblueeyes]{\textbf{@gy:} #1}}
  \renewcommand{\atacracomment}[1]{\todo[size=\footnotesize,color=babyblueeyes]{\textbf{@atab:} #1}}
  \renewcommand{\omcracomment}[1]{\todo[size=\footnotesize,color=red]{\textbf{onur:} #1}}
  \renewcommand{\agycra}[1]{\textcolor{amber}{#1}}

  \renewcommand{\agycra}[1]{\textcolor{blue}{#1}}
\fi

\newif\ifcry
\cryfalse

\newcommand{\agycrytodo}[1]{}

\newcommand{\agycrycomment}[1]{}
\newcommand{\atacrycomment}[1]{}
\newcommand{\omcrycomment}[1]{}
\ifcry
  \renewcommand{\agycrytodo}[1]{\todo[size=\footnotesize,color=red]{\textbf{TODO:} #1}}
  \renewcommand{\agycrycomment}[1]{\todo[size=\footnotesize,color=babyblueeyes]{\textbf{@gy:} #1}}
  \renewcommand{\atacrycomment}[1]{\todo[size=\footnotesize,color=babyblueeyes]{\textbf{@atab:} #1}}
  \renewcommand{\omcrycomment}[1]{\todo[size=\footnotesize,color=red]{\textbf{onur:} #1}}

\fi

\newif\ifcrs
\crsfalse

\newcommand{\agycrstodo}[1]{}

\newcommand{\agycrscomment}[1]{}
\newcommand{\atacrscomment}[1]{}
\newcommand{\omcrscomment}[1]{}
\ifcrs
  \renewcommand{\agycrstodo}[1]{\todo[size=\footnotesize,color=red]{\textbf{TODO:} #1}}
  \renewcommand{\agycrscomment}[1]{\todo[size=\footnotesize,color=babyblueeyes]{\textbf{@gy:} #1}}
  \renewcommand{\atacrscomment}[1]{\todo[size=\footnotesize,color=babyblueeyes]{\textbf{@atab:} #1}}
  \renewcommand{\omcrscomment}[1]{\todo[size=\footnotesize,color=red]{\textbf{onur:} #1}}

\fi

\newif\ifarb
\arbfalse

\newcommand{\agyarbtodo}[1]{}
\newcommand{\agyarb}[1]{#1}

\newcommand{\agyarbcomment}[1]{}
\newcommand{\ataarbcomment}[1]{}
\newcommand{\omarbcomment}[1]{}
\ifarb
  \renewcommand{\agyarbtodo}[1]{\todo[size=\footnotesize,color=red]{\textbf{TODO:} #1}}
  \renewcommand{\agyarbcomment}[1]{\todo[size=\footnotesize,color=babyblueeyes]{\textbf{@gy:} #1}}
  \renewcommand{\ataarbcomment}[1]{\todo[size=\footnotesize,color=babyblueeyes]{\textbf{@atab:} #1}}
  \renewcommand{\omarbcomment}[1]{\todo[size=\footnotesize,color=red]{\textbf{onur:} #1}}
  \renewcommand{\agyarb}[1]{\textcolor{amber}{#1}}

  \renewcommand{\agyarb}[1]{\textcolor{blue}{#1}}
\fi

\ifrevision

  \newcommand{\param}[1]{{#1}}
  
  \newcommand{\atb}[1]{\textcolor{coolblack}{#1}}
  
  \newcommand{\agycomment}[1]{}
  \newcommand{\agytodo}[1]{}
  \newcommand{\loiscomment}[1]{}
  \newcommand{\hluocomment}[1]{}
  \newcommand{\atbc}[1]{}
  \newcommand{\ignore}[1]{}
  
\else

  \ifsubmission   %%% SUBMISSION MACROS

    \newcommand{\param}[1]{#1}
    \newcommand{\atb}[1]{#1}

    \newcommand{\agycomment}[1]{}
    \newcommand{\agytodo}[1]{}
    \newcommand{\loiscomment}[1]{}
    \newcommand{\hluocomment}[1]{}
    \newcommand{\atbc}[1]{#1}
    \newcommand{\ignore}[1]{}

  \else           %%% DRAFT HERE
    \newcommand{\param}[1]{\colorbox{amber}{#1}}

    \newcommand{\atb}[1]{\textcolor{ups-truck}{#1}}
    \newcommand{\atbc}[1]{\textcolor{ups-truck}{\textbf{[@atb: #1]}}}

    \newcommand{\loiscomment}[1]{\noindent{\color{red}{\bf \fbox{Lois: }}{\it#1}}}
    \newcommand{\agycomment}[1]{\textcolor{awesome}{\textbf{[@gy: {\it#1}]}}}
    \newcommand{\hluocomment}[1]{\textcolor{moegi}{\textit{[@hluo: #1]}}}
    \newcommand{\agytodo}[1]{\todo[size=\footnotesize,color=jazzberryjam]{\textbf{@gy:} #1}}
    \newcommand{\ignore}[1]{}
  \fi
\fi

\ifiscarev

\newcommand{\cqlabel}[1]{\todo[size=\small,color=jazzberryjam]{\textbf{\textcolor{white}{#1}}}}
\newcommand{\iqlabel}[1]{\todo[size=\small,color=blue]{\textbf{\textcolor{white}{#1}}}}
\else

\newcommand{\cqlabel}[1]{{}}
\newcommand{\iqlabel}[1]{{}}
\fi

\newcommand{\memacc}[0]{{memory} access}
\newcommand{\memaccs}[0]{\memacc{}es}
\newcommand{\dura}[0]{\gls{hira}}

\newcommand{\doduo}[0]{\dura{}}

\newcommand{\hira}[0]{\dura{}}
\newcommand{\hiraop}[0]{\dura{} operation}

\newacronym{hira}{HiRA}{Hidden Row Activation}
\newacronym{hirasched}{HiRA{-}MC}{the {HiRA Memory Controller}}

\newcommand{\periodicreduction}[0]{\SI{35.4}{\percent}}
\newcommand{\reactivereduction}[0]{\SI{11.4}{\percent}}

\newcommand{\chipcnt}[0]{\param{\SI{56}{\percent}}}

\newcommand{\wlcnt}[0]{\param{{125}}}
\newcommand{\ncores}[0]{\param{8}}

\newcommand{\covref}[0]{\param{\SI{32}{\percent}}}

\newcommand{\hcfirst}[0]{N_{RH}}
\newacronym{hcfirst}{RowHammer threshold}{the minimum \agycru{hammer count} \agycru{required to induce the first RowHammer bit flip}} %at which the first bit error is observed}
\newcommand{\hcdeadline}[0]{N_{RefSlack}}
\newacronym{hcdeadline}{$\hcdeadline{}:t_{RefSlack}/t_{RC}$}{the maximum \agycri{number of activations} that an \agycri{attacker} can \agycri{perform within a \gls{trrslack}}}
\newacronym{ber}{$BER$}{bit error rate}
\newacronym{wcdp}{$WCDP$}{worst-case data pattern}

\newacronym{taggon}{$t_{AggOn}$}{the time that an aggressor row stays active}
\newacronym{taggoff}{$t_{AggOff}$}{the time that an aggressor row stays precharged}
\newacronym{tras}{$t_{RAS}$}{charge restoration latency}
\newacronym{trp}{$t_{RP}$}{precharge latency}
\newcommand{\trc}[0]{t_{RC}}
\newacronym{trc}{$\trc{}$}{row activation cycle}
\newacronym{trcd}{$t_{RCD}$}{row-to-column delay or row activation latency}
\newcommand{\tfaw}[0]{t_{FAW}}
\newacronym{tfaw}{$\tfaw{}$}{four row activation window}
\newcommand{\trefw}[0]{t_{REFW}}
\newacronym{trefw}{$\trefw{}$}{refresh window}
\newcommand{\trefi}[0]{t_{REFI}}
\newacronym{trefi}{$\trefi{}$}{refresh interval}
\newacronym{trrslack}{$t_{RefSlack}$}{{the maximum delay between the time a {periodic}/{preventive} refresh is generated and the time the refresh is performed}}
\newacronym{tapa}{$t_{APA}$}{the latency of issuing $ACT-PRE-ACT$ command sequence}
\newacronym{ref}{$REF$}{refresh}
\newacronym{act}{$ACT$}{activate}
\newacronym{pre}{$PRE$}{precharge}
\newcommand{\trfc}[0]{t_{RFC}}
\newacronym{trfc}{$\trfc{}$}{refresh latency}
\newacronym{iqr}{$IQR$}{interquartile range}
\newacronym{cv}{$CV$}{the coefficient of variation}
\newacronym{hc}{$HC$}{hammer count}
\newcommand{\pth}[0]{p_{th}}
\newacronym{pth}{$\pth{}$}{\agycra{PARA's probability threshold}}
\newcommand{\pf}[0]{p_{failure}}
\newacronym{pf}{$\pf{}$}{failure probability over a sufficiently long time}
\newcommand{\prh}[0]{p_{RH}}
\newacronym{prh}{$\prh{}$}{reliability target for a \gls{trefw}}

\newcommand{\cchip}[0]{D_{chip}}
\newacronym{cchip}{$\cchip{}$}{chip density}

\newcommand{\rbcpki}[0]{RBCPKI}
\newacronym{rbcpki}{$\rbcpki{}$}{row buffer conflicts per kilo instruction}

\newcommand{\mpki}[0]{MPKI}
\newacronym{mpki}{$\mpki{}$}{misses per kilo instruction}

\newcommand{\vdd}[0]{V_{DD}}
\newacronym{vdd}{$\vdd{}$}{supply voltage}

\newcommand{\gnd}[0]{GND}
\newacronym{gnd}{$\gnd{}$}{ground}

\newcommand{\rd}[0]{RD}
\newacronym{rd}{$\rd{}$}{read}

\newcommand{\dramwr}[0]{WR}
\newacronym{wr}{$\dramwr{}$}{write}

\newcommand{\equ}[2]{
\vspace{-0.75em}
\begin{equation}
    \footnotesize
    #1
    \label{#2}
    \vspace{0.25em}
\end{equation}
}

\newcommand{\footref}[1]{\textsuperscript{\ref{#1}}}
\newcommand{\secref}[1]{Section~\ref{#1}}
\newcommand{\tabref}[1]{Table~\ref{#1}}
\newcommand{\algref}[1]{Algorithm~\ref{#1}}
\newcommand{\figref}[1]{Figure~\ref{#1}}

\newcommand{\expref}[1]{Expression~\ref{#1}}
\newcommand{\expsref}[1]{Expressions~\ref{#1}}
\renewcommand{\figref}[1]{Fig.~\ref{#1}}

\renewcommand{\expref}[1]{Exp.~\ref{#1}}
\renewcommand{\expsref}[1]{Exps.~\ref{#1}}
\renewcommand{\figurename}{Fig.}
\renewcommand{\secref}[1]{§\ref{#1}}

\newcommand{\rhdefinition}[0]{\cite{kim2014flipping, kim2020revisiting, mutlu2017rowhammer, yang2019trap, mutlu2019rowhammer, walker2021ondramrowhammer, park2016experiments, park2016statistical, yaglikci2022understanding}}
\newcommand{\rhdefrefresh}[0]{\cite{AppleRefInc, kim2014flipping, kim2014architectural, bains2015row, aweke2016anvil, bains2016row, bains2016distributed, son2017making, seyedzadeh2018cbt, you2019mrloc, lee2019twice, park2020graphene, yaglikci2021security, frigo2020trrespass, kang2020cattwo, hassan2021utrr, qureshi2022hydra, kim2022mithril, devaux2021method, lee2021cryoguard, marazzi2022protrr, zhang2022softtrr, joardar2022learning}}
\newcommand{\rhdef}[0]{\cite{AppleRefInc, kim2014flipping, kim2014architectural, aweke2016anvil, bains2015row, bains2016row, bains2016distributed, son2017making, seyedzadeh2018cbt, you2019mrloc, lee2019twice, park2020graphene, yaglikci2021security, yaglikci2021blockhammer, frigo2020trrespass, kang2020cattwo, hassan2021utrr, qureshi2022hydra, saileshwar2022randomized, brasser2017can, konoth2018zebram, van2018guardion, greenfield2012throttling, kim2022mithril, lee2021cryoguard, marazzi2022protrr, zhang2022softtrr, joardar2022learning, juffinger2023csi}}
\newcommand{\perrefreshscaling}[0]{\cite{nguyen2018nonblocking, liu2012raidr, baumann2005radiation, liu2013experimental, chang2014improving, kang2014co, mukundan2013understanding, nair2014refresh, nair2013arch}}
\newcommand{\rhsafe}[0]{\cite{frigo2020trrespass, lee2014green, micron2014ddr4}}
\newcommand{\rhworse}[0]{\cite{kim2014flipping,mutlu2017rowhammer,mutlu2019rowhammer,kim2020revisiting, frigo2020trrespass, hassan2021utrr, orosa2021deeper, jattke2022blacksmith, yaglikci2022understanding, deridder2021smash}}
\newcommand{\rhdefworse}[0]{\cite{kim2020revisiting,hassan2021utrr, orosa2021deeper, yaglikci2021security, yaglikci2021blockhammer, park2020graphene, qureshi2022hydra}}
\newcommand{\rhattacks}[0]{\cite{seaborn2015exploiting, van2016drammer, gruss2016rowhammer, razavi2016flip, pessl2016drama, xiao2016one, bosman2016dedup, bhattacharya2016curious, qiao2016new, jang2017sgx, aga2017good, mutlu2017rowhammer, tatar2018defeating, gruss2018another, lipp2018nethammer, van2018guardion, frigo2018grand, cojocar2019eccploit,  ji2019pinpoint, mutlu2019rowhammer, hong2019terminal, kwong2020rambleed, frigo2020trrespass, cojocar2020rowhammer, weissman2020jackhammer, zhang2020pthammer, rowhammergithub, yao2020deephammer,deridder2021smash,
hassan2021utrr, jattke2022blacksmith, tol2022toward,burleson2016invited, brasser2017can, kim2014flipping, kogler2022half}}
\newcommand{\salprefs}[0]{\cite{salp, chang2014improving, wang2020figaro, zhang2014cream}}
\begin{document}

\maketitle
\thispagestyle{firstpage}
\pagestyle{plain}
\glsresetall
\begin{abstract}
{DRAM is the building block of modern main memory systems. DRAM cells must be periodically refreshed to prevent data loss. Refresh operations degrade system performance by interfering with memory accesses.}
{{As} DRAM chip density increases with technology node scaling, refresh operations {also increase because}:}
1)~{the number of DRAM rows in a chip increases}; and 2)~DRAM cells {need additional refresh operations to mitigate bit failures caused by RowHammer, a {failure} mechanism that {becomes worse with technology node scaling}.} 
{Thus,} it is critical to {enable} refresh {operations} at low performance overhead.
{To {this end},} we propose {a new operation,} \gls{hira}, {and
{\gls{hirasched} to perform \gls{hira} operations.}
} 

{\gls{hira} hides a refresh operation's latency by refreshing a row {concurrently} with accessing or refreshing another row within the same bank. Unlike prior works, \gls{hira} achieves this parallelism without any modifications to off-the-shelf DRAM chips. To do so, it leverages}
{the new observation that two rows in the same bank can be activated without data loss if the rows are connected to different charge restoration circuitry.}
We experimentally demonstrate on \chipcnt{}
real off-the-shelf DRAM chips that {\gls{hira} {can} reliably parallelize} a DRAM row's {refresh operation} {with refresh or {activation of} {any of the}} \covref{} of the rows {within the same bank. By doing so,}
\gls{hira} reduces {the overall latency of two} refresh {operations} by \SI{51.4}{\percent}.

\gls{hirasched} modifies the memory request scheduler to perform \gls{hira} when a refresh operation can be {performed concurrently} with a \memacc{} or another refresh.
{Our system-level evaluations show that}
{HiRA-MC {increases system performance by \SI{12.6}{\percent} and $3.73\times$ {as it reduces}} the performance degradation {due to} periodic refreshes and refreshes for RowHammer {protection} (preventive refreshes){, respectively,} for future DRAM chips with increased density and RowHammer vulnerability.
}
\end{abstract}
\glsresetall
\setstretch{0.968}
\section{Introduction}
\label{sec:introduction}
DRAM~{\cite{dennard1968dram}} is the prevalent {main memory} technology used in a wide variety of computing systems from cloud servers to mobile devices {due to its high density and low latency}.
{A DRAM cell encodes a bit of data as electrical charge, {which} inherently leaks over time~\cite{keeth2001dram}. Therefore,}
to ensure reliable operation and data integrity, a DRAM cell needs to be periodically refreshed~\cite{jedecddr, jedec2017ddr4, jedec2020ddr5, jedec2015lpddr4, jedec2020lpddr5, jedec2015hbm}. 
Unfortunately, these refresh operations degrade system performance by interfering with memory accesses~\cite{liu2012raidr, chang2014improving}. {{During a} refresh operation{, which} is performed at row granularity (e.g., 8KB), the memory cannot service any {requests to} the DRAM bank~(e.g.,~512MB) or rank~(e.g.,~8GB) that contains the refreshed row~\cite{micron2014ddr4, jedec2017ddr4, jedec2015lpddr4}.}

{As DRAM density increases with technology node scaling, the performance overhead of refresh also increases due to three major reasons.}
{First, as the {DRAM} chip density increases, more DRAM {rows} need to be {periodically} refreshed in a DRAM chip~\cite{jedecddr, jedec2017ddr4, jedec2020ddr5, jedec2015lpddr4, jedec2020lpddr5, jedec2015hbm}.
{Second, as {DRAM} technology node scales down, DRAM cells become smaller and thus can store less amount of charge, requiring them to be refreshed more frequently~{\perrefreshscaling{}}.}
{Third}, with increasing {DRAM} density, DRAM cells are placed closer to each other, exacerbating charge leakage via a disturbance error mechanism called RowHammer~\rhdefinition{}, and thus requiring additional refresh operations {(called \emph{preventive} refreshes)} to avoid data {corruption} due to RowHammer~\rhdefrefresh{}.}
Prior work shows that 1)~{RowHammer} can be exploited to escalate privilege, leak private data, and manipulate critical application outputs~\rhattacks{}; and 2)~modern DRAM chips, including the ones that are marketed as RowHammer-safe~\rhsafe{}, are more vulnerable to RowHammer than their predecessors~\rhworse{}. {Therefore, {defending against} RowHammer is critical for secure system operation and {doing so likely} requires aggressively refreshing the cells disturbed by RowHammer~\rhdefworse{}.
As a result of {these three major reasons, newer generations of} DRAM chips require performing more refresh operations compared to their predecessors{. T}hus{,} it is critical to reduce the performance overhead of refresh{es}.}

Prior works suggest reducing refresh latency by 1)~accelerating the charge restoration process~\cite{luo2020clrdram, hassan2019crow} and 2)~exploiting parallelism across subarrays within a DRAM bank~\salprefs{}. Unfortunately, these proposals require modifications to DRAM circuitry, making them {un}suitable for already deployed off-the-shelf DRAM chips.
{Therefore,} it is {important} to find alternative solutions to reduce the negative performance impact of {a} refresh operation {with no modifi{c}ations to the DRAM chip circuitry}. 

\textbf{Our goal} is to reduce the refresh latency in off-the-shelf DRAM chips with \emph{no} modifications to DRAM circuitry. 
To this end, we propose {a new operation called} \gls{hira} {and {\gls{hirasched}}} to perform \gls{hira} operations.

\gls{hira} enables refreshing a DRAM row while refreshing or accessing another DRAM row within the same bank.
\gls{hira} leverages the new observation that opening two rows, {whose charge restoration circuitries are electrically isolated from each other{,} {in rapid succession}, allows refreshing one row while refreshing or accessing the other row.} 
{To {open two such rows in rapid succession},} \gls{hira} {uses a carefully{-}engineered sequence of} \gls{act} and \gls{pre} commands, already implemented in off-the-shelf DRAM chips for opening and closing DRAM rows, respectively.
We experimentally demonstrate on \chipcnt{} real off-the-shelf DRAM chips that \gls{hira} 1)~{reliably parallelizes a DRAM row's {refresh operation} {with refresh or {activation} of any of the} \covref{} of the rows within the same bank}
{and 2)~}effectively reduces the {overall} latency of refreshing two rows
by \param{\SI{51.4}{\percent}}.

{\Glsfirst{hirasched} leverages {the} \gls{hira} {operation} {to improve}} system {performance by performing} \emph{two} main {tasks}. First, {it} queues {each} refresh request {with a time slack before the refresh needs to be performed and assigns the refresh request} a deadline.
Second, {it} observes the \memaccs{} {at} real-time to find a \memacc{} that can be {performed concurrently} with a {queued} refresh {request}.
\gls{hirasched} ensures that each periodic and preventive refresh request is performed by its deadline. \gls{hirasched} achieves this by taking one of three possible actions, in {the following} order: 1)~refresh {a DRAM row} {concurrently} with a \memacc{} {(refresh-access parallelization)} before the refresh operation's deadline; 2)~refresh {a row} {concurrently} with another refresh operation {(refresh-refresh parallelization)} if {refresh-access parallelization is \emph{not} possible} until the refresh operation's deadline; or 3)~perform a refresh operation {right} at its deadline if neither 
{refresh-access nor refresh-refresh parallelization is possible.}
We evaluate \gls{hirasched}'s hardware complexity and show that it consumes only \param{{\SI{0.00923}{\milli\meter\squared}}} chip area and {responds to queries within {{\SI{6.31}{\nano\second}}} (in parallel to a \gls{pre} command with a latency of \SI{14.5}{\nano\second})}{.}
To evaluate \gls{hirasched}'s performance benefits, we conduct cycle-level simulations on \wlcnt{} multiprogrammed workloads.
Our analysis shows that {1)~\emph{without} \gls{hira}, periodic and preventive refresh operations cause \SI{26.3}{\percent} and \SI{96.0}{\percent} performance overhead for future DRAM chips with high density and high RowHammer vulnerability, respectively}, compared to an ideal memory controller that does \emph{not} perform periodic or preventive refreshes {and 2)~\gls{hirasched} {increases system performance by \SI{12.6}{\percent} and $3.73\times$ {as it reduces}} the performance degradation {due to} periodic refreshes and preventive refreshes{, respectively,} for future DRAM chips with increased density and RowHammer vulnerability.}

This paper makes the following contributions: 
\glsresetall
\squishlist
\item This is the first {work to} show that refresh-refresh and refresh-access parallelization within a bank is possible {in} off-the-shelf DRAM chips by issuing a carefull{-}engineered sequence of \gls{act} and \gls{pre} {commands}, which we call \gls{hira}.  
\item We experimentally demonstrate on \chipcnt{} real {DDR4} DRAM chips that {\gls{hira}} 1)~reduces the latency of refreshing two rows back-to-back by \param{\SI{51.4}{\percent}}, 
and 2)~{reliably parallelizes a DRAM row's {refresh operation} {with refresh or {activation} of any of the} \covref{}}
of the rows
in the same bank.
\item We {design} {\gls{hirasched} to perform \gls{hira} operations}. We show that {\gls{hirasched} significantly {improves system performance by \SI{12.6}{\percent} and $3.73\times$ {as it reduces}} the performance degradation {due to} periodic refreshes and {preventive} refreshes for RowHammer}{,}
respectively. 
 
\squishend

\glsresetall
\setstretch{0.96}
\section{Background}
\label{sec:background}

This section describes the background required to understand the rest of the paper. For a more comprehensive description of DRAM organization and operation, we refer the reader to~\cite{salp,liu2012raidr,lee2013tiered,keeth2001dram, lee2015decoupled}.
\subsection{DRAM {O}rganization}
DRAM is organized hierarchically. The {memory controller} accesses DRAM modules {via a} memory channel.
{A} DRAM module {has} one or more ranks{, each of which contains multiple DRAM chips} that work in lock-step. 
{{Each DRAM chip consists of} multiple DRAM banks{, {which} shar{e} the I/O circuitry {of the chip (called \emph{{chip I/O}})}}.}

\begin{figure}[h]
    \centering
    \includegraphics[width=\linewidth]{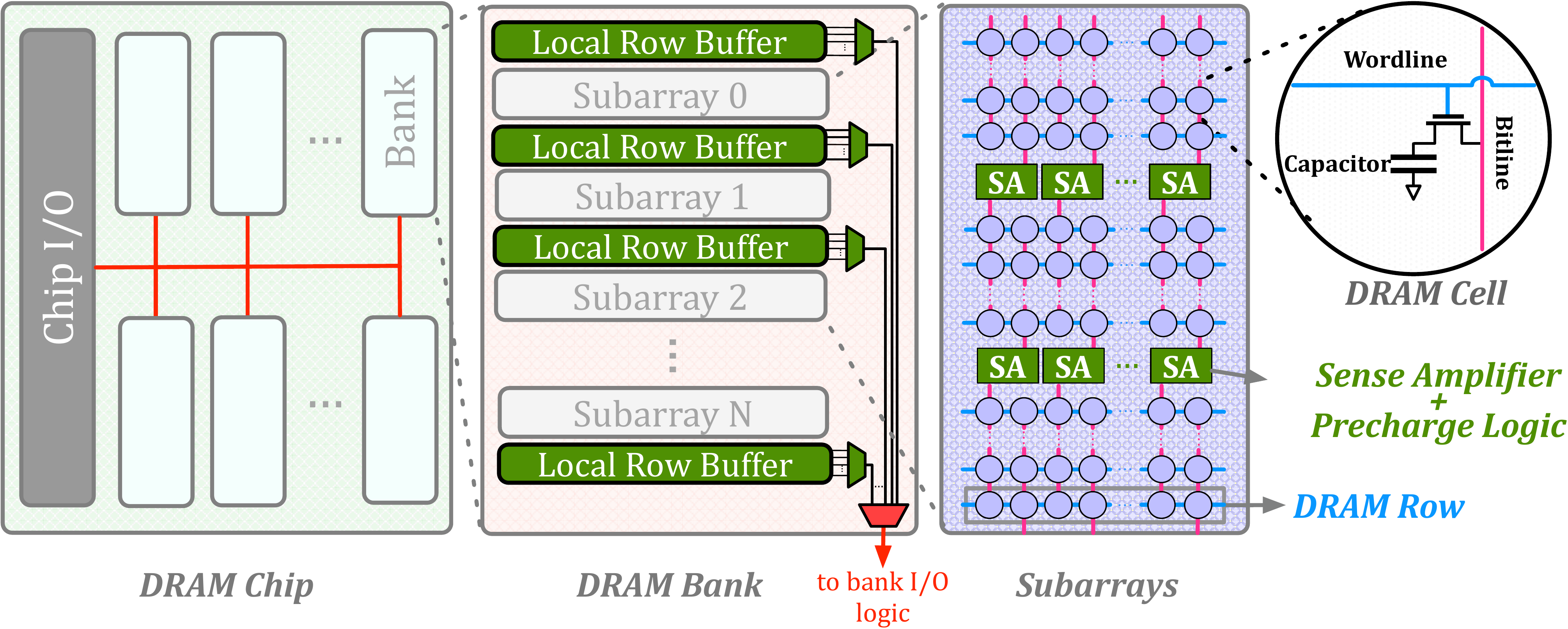}
    \caption{DRAM organization}
    \label{fig:DRAM_organization}
\end{figure}

\noindent\textbf{Bank Organization.} \figref{fig:DRAM_organization} shows a typical DRAM chip organization, {containing multiple banks.}
A DRAM bank is composed of multiple subarrays{, {which} shar{e} the I/O circuitry of the bank {(called \emph{{bank I/O}})}}. Each subarray contains a two{-}dimensional array of DRAM cells, organized as rows and columns {and a local row buffer}. A DRAM cell stores data as {electrical} charge in a capacitor, which is {accessed} via an access transistor.
{The gate of an access transistor is driven by a row-wide {wire}, called wordline. {T}he access transistor connects the cell capacitor to a column-wide structure, called bitline. {A} bitline is connected to a \emph{sense amplifier (SA)} and \emph{precharge} circuitry.}

\noindent\textbf{Open-bitline Architecture.} To optimize the size and layout of subarrays, the common open-bitline architecture~\cite{chang2016low,keeth2007dram, luo2020clrdram} places the SAs {on} both {ends} (i.e., top and bottom) of the subarray (as depicted in \figref{fig:DRAM_organization}). In this architecture, {horizontally} adjacent DRAM cells are connected {via} bitlines to SAs {on both ends} of the subarray.

\subsection{DRAM Operation}
The memory controller {implements a memory request scheduler to issue} DDRx commands. We describe the four DDRx commands that are relevant to this work.

\noindent
{\textbf{Precharge.}} {A bitline needs to be precharged to half of {the} supply voltage ($V_{DD}/2$) before accessing a DRAM row {in a subarray}. To do so,} the \gls{pre} command prepares {the subarray} {for accessing data in DRAM cells} {using the precharge logic placed next to the sense amplifiers}.

\noindent
{\textbf{Row Activation.} To access a DRAM cell, the memory controller needs to open the DRAM row containing the cell. To do so, the memory controller issues an \gls{act} command.
An \gls{act} command is performed in \param{{four}} steps.
First, it drives {the target} wordline with a high voltage {level} to turn on the access transistors within the DRAM row {(i.e., open the DRAM row)}.
Second, turning on {an access transistor} initiates {the} \emph{charge sharing} process between the DRAM cell and its bitline.
Third, the charge sharing process causes a small deviation {in} the bitline voltage.
Fourth, {the sense amplifier is enabled to sense the voltage deviation on the bitline and amplify the bitline voltage to {the} level of {either the} \gls{vdd} or \gls{gnd} (depending on the voltage deviation {on the bitline}).} {In doing so, an \gls{act} command copies the data in the open DRAM row to the local row buffer.}

\noindent
{\textbf{Column Access.}}
{Once the open row is copied to the local row buffer, it can be accessed by \gls{rd} or {modified} by \gls{wr} column access commands in the local row buffer using the bank I/O {circuitry}.}
{Only one DRAM row in {a bank} can be open at a given time~\cite{jedec2017ddr4, salp}.}

\noindent
{\textbf{Timing Parameters.}}
{DRAM specifications (e.g., DDR4~\cite{jedec2017ddr4}) define timing parameters that the memory controller obeys while scheduling DRAM commands. {Four timing parameters are important to understand the rest of the paper. First,} consecutive row activation and column access commands (e.g., \gls{act} to \gls{rd}) must be separated in time by at least the \gls{trcd}. \gls{trcd} ensures that the deviation {i}n the bitline voltage exceeds the reliable sensing threshold of the sense amplifier during row activation~\cite{kim2019d,chang2016understanding}. {Second,} consecutive {\gls{act} and \gls{pre}} commands must be separated by at least the \gls{tras}. \gls{tras} ensures that the charge levels of all DRAM cells in the open row are fully restored before the row is closed. {Third,} consecutive precharge and row activation commands must be separated by at least the \gls{trp}. \gls{trp} ensures that the bitline is fully precharged to $V_{DD}/2$, so that the {next row activation} can be reliably performed{.}
{Fourth, {the} time window between two consecutive row activations targeting a DRAM bank, {i.e.,} \gls{trc} {must be at least} as large as the sum of \gls{tras} and \gls{trp}.}}}

\subsection{DRAM Refresh} 

{C}harge stored in the capacitor of a DRAM cell {leaks} over time. {Therefore, the charge in the cell's capacitor must be periodically restored (i.e., the DRAM cell must be \emph{refreshed}) to maintain data integrity.}
The time interval during which a cell can retain its charge without being refreshed is called the {cell's} \emph{retention time}. {A DRAM cell needs to be refreshed once every \gls{trefw}, which is typically \SI{64}{\milli\second} (e.g., in DDR4~\cite{jedec2017ddr4}) or \SI{32}{\milli\second} (e.g., in DDR5~\cite{jedec2020ddr5}).}

{To perform refresh operations,} the memory controller {periodically issues} a \gls{ref} command {to a DRAM rank} {{after every} \gls{trefi} (e.g., \SI{7.8}{\micro\second}~\cite{jedec2017ddr4} or {\SI{3.9}{\micro\second}~\cite{jedec2020ddr5}}).}
Each \gls{ref} command refreshes a {number} of rows in the DRAM chip {based on the chip's capacity. T}he DRAM chip
{internally decides} which rows {and how many rows} to refresh{, but does \emph{not} expose this information to the memory controller.} 

{Issuing a \gls{ref} command {makes the} DRAM rank unavailable for {a time window called \gls{trfc}, during which the rank {\emph{cannot}} receive any commands.}} {{Unfortunately,} \gls{trfc} needs to be large enough {(e.g., \SI{350}{\nano\second}~\cite{jedec2017ddr4})} such that multiple rows can be refreshed with a \gls{ref} command.} {Thus, {issuing a \gls{ref} command can} increase {the access} latency {of memory requests} and cause system-wide slowdown.} With increasing DRAM chip density, more DRAM rows need to be refreshed, exacerbating the negative performance impact of DRAM refresh~\cite{nguyen2018nonblocking, liu2012raidr}. 

\subsection{RowHammer}
\label{sec:background_rowhammer}
Modern DRAM devices suffer from disturbance errors that happen when a DRAM row {(the aggressor row)} is repeatedly and rapidly activated~\cite{kim2014flipping, mutlu2017rowhammer, mutlu2019rowhammer}. These disturbance errors manifest {in DRAM rows neighboring the aggressor row (i.e., victim rows)} after the aggressor row's activation count {(i.e., hammer count)} reaches a certain threshold value within a refresh window, which we call {the \emph{RowHammer threshold ($N_{RH}$)}}~\cite{kim2020revisiting, kim2014flipping, orosa2021deeper,yaglikci2022understanding}. {As} DRAM cells become smaller and closer to each other {with technology node scaling}, RowHammer vulnerability {becomes worse}~\rhworse{}.
Given the severity of {the} RowHammer {vulnerability}, {many prior works propose refreshing the potential victim rows to prevent RowHammer bit flips, which we call \emph{preventive refresh}~\rhdefrefresh{}}.

% \pagebreak
\section{HiRA: Hidden Row Activation}
\label{sec:doduo}

\begin{figure*}[!ht]
    \centering
        \includegraphics[width=0.7\linewidth]{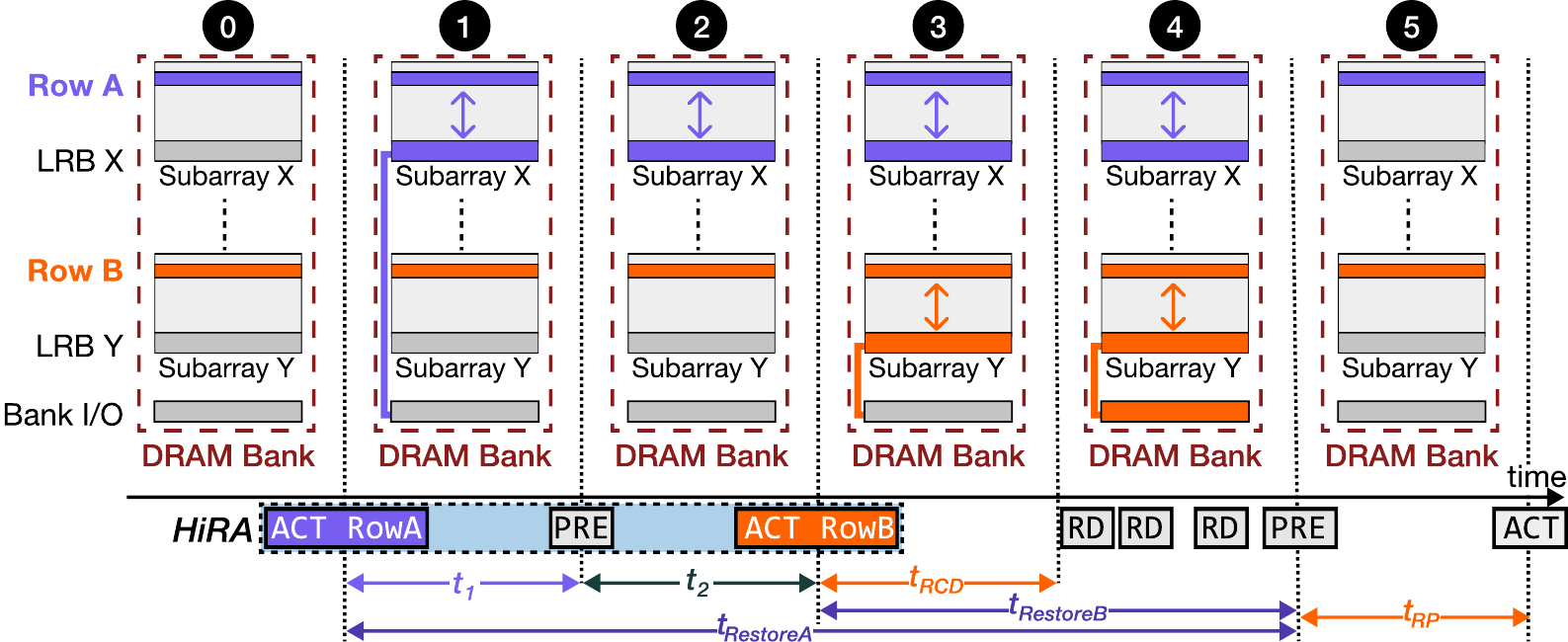}
    \caption{Performing a \doduo{} operation and its effects on a DRAM bank. Command timings are not {to scale}. {LRB: Local Row Buffer}}
    \label{fig:cmdseq}
\end{figure*}

\noindent\textbf{{Overview}.}
{We develop the \gls{hira} operation} for concurrently activating two DRAM rows within a {DRAM} bank. 
\gls{hira} overlaps the latency of refreshing a DRAM row with the latency of refreshing or 
{activating} 
another DRAM row in the same DRAM bank.
\figref{fig:cmdseq} demonstrates how {a} \gls{hira} {operation} is performed by issuing a carefully{-}engineered sequence of $ACT~RowA$, \gls{pre}{, and $ACT~RowB$} commands with two customized timing parameters: $t_1$ ({$ACT~RowA$} to $PRE$ latency) and $t_2$ ($PRE$ to {$ACT~RowB$} latency). {A \hiraop{}'s first \gls{act} refreshes $RowA$ and the second \gls{act} refreshes $RowB$ {and} opens it for column accesses. Since \gls{act} and \gls{pre} commands} {are already implemented in off-the-shelf DRAM chips{,}}  \hira{} \emph{does not} require modifications to the DRAM chip circuitry.

{At a high level, a \hiraop{} 1)~activates $RowA$, 2)~precharges the bank \emph{without} disconnecting $RowA$ from its local row buffer, and 3)~activates $RowB$. In doing so, it allows the memory controller to 1)~perform two refresh operations on $RowA$ and $RowB$ with a latency significantly smaller than two times the \gls{trc} (i.e., refresh-refresh parallelization) and~2)~{activate} $RowB$ for column accesses (i.e., only $RowB$'s local row buffer gets connected to the bank I/O after performing a \hiraop{}) {{concurrently} with refreshing $RowA$} (i.e., refresh-access parallelization).}

\noindent\textbf{\gls{hira} Operation Walk-Through.} \figref{fig:cmdseq} demonstrates how a \doduo{} operation is performed and how it affects the state of a DRAM bank.
Initially (\circled{0}) the DRAM bank {is in} precharged {state and thus there is no active row}. \doduo{} begins {by} issuing an {\gls{act}} command targeting $RowA$, which connects $RowA$'s cells to \emph{local row buffer X}~(\circled{1}). Then, a precharge command is issued to disconnect \emph{local row buffer {(LRB)} X} {from the} \emph{{bank I/O}}~(\circled{2}). This precharge operation is interrupted by issuing a new row activation, targeting $RowB$ {in a completely separate subarray Y}~(\circled{3}), to avoid breaking the connection between the \emph{local row buffer X} and $RowA$. {Therefore, the sense amplifiers in the local row buffer X continue charge restoration of $RowA$. Thus, \gls{hira} overlaps the latency of refreshing $RowA$ with the latency of activating $RowB$.} It is important that the subarray that contains $RowB$ (\emph{subarray~Y}) is physically isolated from the subarray that contains $RowA$ (\emph{subarray~X}), such that subarrays X and Y do not share any bitline or sense amplifier and thus activating $RowB$ does not {affect} the voltage levels on subarray~X's bitlines~(\circled{3}). The {\hiraop{}} completes when the second row activation is issued, {after which both $RowA$ and $RowB$ are connected to their local row buffers without corrupting each other's data~(\circled{3})}. Following a \doduo{} operation, $RowB$'s content can be read by issuing {\gls{rd}} commands once \gls{trcd} is satisfied~(\circled{4}). To close both $RowA$ and $RowB$, issuing one precharge command is enough~(\circled{5}).\footnote{{Our experiments verify that {issuing one precharge command is enough} {to {reliably} close \emph{both} rows} in {all} \chipcnt{} real DRAM chips {we test}. We hypothesize that issuing a \gls{pre} command disables all wordlines and precharges all bitlines in a DRAM bank because {the precharge command} is \emph{not} provided with a row address~\cite{jedecddr, jedec2015hbm, jedec2017ddr4, jedec2015lpddr4, jedec2020lpddr5, jedec2020ddr5, micron2014ddr4}.}}

\noindent\textbf{Charge Restoration after \gls{hira}.}
\figref{fig:cmdseq} highlights the charge restoration time that $RowA$ and $RowB$ experience as $t_{RestoreA}$ and $t_{RestoreB}$, respectively.
To ensure charge restoration {happens correctly} for $RowA$ and $RowB$, both $t_{RestoreA}$ and $t_{RestoreB}$ should be larger than or equal to the {existing} \gls{tras} {timing parameter}~\cite{jedecddr, jedec2015hbm, jedec2017ddr4, jedec2015lpddr4, jedec2020lpddr5, jedec2020ddr5, micron2014ddr4}. Because we do \emph{not} modify the timing constraints of the second \gls{pre} command (\circled{5}), {existing DRAM} timing restrictions already ensure that $t_{RestoreB}$ is larger than or equal to the nominal \gls{tras}. Since $t_{RestoreA}$ is already larger than $t_{RestoreB}$ {(because $RowA$ is activated earlier than $RowB$)}, we conclude that {charge restoration happens correctly for} both rows.

\noindent\textbf{\doduo{}'s Novelty.} \doduo{}'s command sequence ($ACT$-$PRE$-$ACT$) is similar to the command sequences used in {multiple} prior works~\cite{gao2019computedram, olgun2021quactrng, olgun2021pidram}. {These prior works use the $ACT$-$PRE$-$ACT$ command sequence to activate two rows in the \emph{same} subarray for various purposes {(which we explain below)}. In contrast, \doduo{}'s purpose is to activate two rows in \emph{different} subarrays such that we can
{refresh a DRAM row {concurrently} with refreshing or activating another row in the same bank.}}

First, ComputeDRAM~\cite{gao2019computedram} {and PiDRAM~\cite{olgun2021pidram}} perform {an} $ACT$-$PRE$-$ACT$ command sequence to enable bulk data copy across DRAM rows {in the same subarray} (also known as RowClone~\cite{seshadri2013rowclone}) in {off-the-shelf} DRAM chips. 
Second, QUAC-TRNG~\cite{olgun2021quactrng} uses $ACT$-$PRE$-$ACT$ command sequence for performing an operation called \emph{{quadruple row activation}}, which concurrently activates four rows whose addresses vary only in the least significant two bits. 
{1)~}{The} RowClone~\cite{seshadri2013rowclone} implementation{s} of {both} {ComputeDRAM~\cite{gao2019computedram} and PiDRAM~\cite{olgun2021pidram}} and {2)~}QUAC-TRNG's~\cite{olgun2021quactrng} quadruple row activation require using two rows within the {\emph{same}} subarray, so that the bitlines and local sense amplifiers are used for sharing the electrical charge across activated DRAM rows. Therefore, these works do \emph{not} {activate} DRAM rows {in \emph{different} subarrays.} 
In contrast, \doduo{} {exclusively} targets two {rows in different subarrays}, so that it enables the memory controller to perform two key operations that were \emph{not} {known to be} possible before {on off-the-shelf DRAM chips}: 1)~concurrently refreshing two rows, and 2)~refreshing one row while {activating another row in a different subarray.}

\noindent\textbf{HiRA's Main Benefit.} \gls{hira} largely overlaps {a DRAM row's charge restoration latency ($t_{RestoreA}$} in \figref{fig:cmdseq}) with the {latency of another row's activation and charge restoration (\gls{trcd} and $t_{RestoreB}$ in \figref{fig:cmdseq}, respectively)}. {Doing so allows \gls{hira} to reduce the latency {of two operations. First, \gls{hira} reduces the latency of} a memory {access} request that is scheduled immediately after a refresh operation. {With \gls{hira}, such {a} request experiences a latency of $t_1+t_2$ (\circled{1} and \circled{2} in \figref{fig:cmdseq}), which can be as small as \SI{6}{\nano\second} (\secref{sec:experiments_coverage}), instead of the nominal row cycle {time} of} \SI{46.25}{\nano\second} (\gls{trc}~\cite{jedec2017ddr4, micron2014ddr4}). {Second, \gls{hira} reduces the overall latency of} refreshing two DRAM rows in the same bank. {With \gls{hira}, such {an} operation takes \emph{only} \SI{38}{\nano\second} ({{{\SI{6}{\nano\second} for {the} \hiraop{} {to complete} (\secref{sec:experiments_coverage}) and \SI{32}{\nano\second} to ensure that $t_{RestoreB}$ is large enough to complete charge restoration~\cite{jedec2017ddr4, micron2014ddr4}}}) instead of}} {the nominal latency of} \SI{78.25}{\nano\second}.{\footnote{{To refresh two rows using nominal timing parameters, {a conventional} memory controller 1)~activates the first row and waits until charge restoration is complete ($t_{RAS}=32ns$), 2)~precharges the bank and waits until all bitlines are ready for the next row activation ($t_{RP}=14.25ns$), and~3)~activates the second row and waits until charge restoration is complete ($t_{RAS}=32ns$)~\cite{jedec2017ddr4, micron2014ddr4}.\label{fn:reftworows}}}}}

\noindent\textbf{\gls{hira} Operating Conditions.}
{A} \gls{hira} operation works {reliably} if {four} conditions are satisfied.
{First,} $t_1$ should be large enough {so that the sense amplifiers are enabled before the {precharge} command is issued {($PRE$ in \figref{fig:cmdseq})}.} 
{Second,}~$t_2$ should be small enough {so that the second activate command ($ACT~RowB$ in \figref{fig:cmdseq}) interrupts the precharge operation \emph{before} $RowA$'s wordline is disabled, allowing charge restoration on $RowA$ to complete correctly.} 
{Third,}~$t_2$ should be large enough to disconnect the local row {buffer X from {the} bank I/O logic if \gls{hira} is performed for refresh-access parallelization{, so that {future} column accesses are performed {only on $RowB$ (LRB~Y).}} This constraint does not apply to refresh-refresh parallelization because the bank I/O logic is not used during refresh.
Fourth,}~$RowA$ and $RowB$ should be located in two different subarrays that are physically isolated from each other, such that they do not share any sense amplifier or bitline.  

\setstretch{0.95}
\section{HiRA in {Off-the-Shelf} DRAM Chips}
\label{sec:characterization}

{In this section, we demonstrate that \gls{hira} works {reliably} on \chipcnt{} real DDR4 DRAM chips.
{Table~\ref{tab:dram_chip_list} provides {the} {chip} density, die revision {(Die Rev.)}, chip organization {(Org.)}, and manufacturing date of tested DRAM modules {where DRAM chips are manufactured by SK Hynix}.{\footnote{{We observe that \gls{hira} reliably works only in DRAM chips from SK Hynix (similar to QUAC-TRNG~\cite{olgun2021quactrng}) out of {40, 40, and 56} DRAM chips that we test from three major {DRAM} manufacturers: Micron, Samsung, and SK Hynix{, respectively}. {\secref{sec:limitations} discusses why we do \emph{not} observe a successful HiRA operation in DRAM chips manufactured by Micron and Samsung.}
{A~is} F4-2400C17S-8GNT from GSKill~\cite{datasheetF42400C17S8GNT}, {B is} KSM32RD8/16HDR from Kingston~\cite{datasheetksm32rd8}, and {C is} HMAA4GU6AJR8N-XN from SK Hynix~\cite{datasheetHMAA4GU6AJR8N}{.}
}}} We report the manufacturing date of these modules in the form of $week-year$.}

\begin{table}[h]
    \caption{Summary of the tested DDR4 DRAM chips {and {key} experimental results}}
    \centering
    \footnotesize{}
    \setlength\tabcolsep{3pt} 
    \begin{tabular}{l|l|cccc|cc}
        
        \toprule
            {\textbf{Model}} & 
            \begin{tabular}[l]{@{}l@{}}\textbf{DIMM} \textbf{Mfr.}\end{tabular}& \begin{tabular}[c]{@{}c@{}}\textbf{Chip}\\\textbf{\textbf{Capacity}}\end{tabular}& \begin{tabular}[c]{@{}c@{}}\textbf{Die}\\\textbf{Rev.}\end{tabular}& \begin{tabular}[c]{@{}c@{}}\textbf{Chip}\\\textbf{Org.}\end{tabular}& \begin{tabular}[c]{@{}c@{}}\textbf{Mfr.}\\\textbf{Date}\end{tabular}& \begin{tabular}[c]{@{}c@{}}\textbf{HiRA}\\\textbf{Cov.$^*$}\end{tabular}& \begin{tabular}[c]{@{}c@{}}\textbf{Norm.}\\$\boldsymbol{N_{RH}^{**}}$\end{tabular}\\
        \midrule
        A0 & \multirow{2}{*}{GSKill~\cite{datasheetF42400C17S8GNT}} & \multirow{2}{*}{4Gb}&\multirow{2}{*}{B}&\multirow{2}{*}{$\times$8}&\multirow{2}{*}{42--20} &\SI{25.0}{\percent}&1.90\\
        A1 & & & & & &\SI{26.6}{\percent}&1.94\\
        \midrule
        B0 & \multirow{2}{*}{Kingston~\cite{datasheetksm32rd8}} & \multirow{2}{*}{8Gb}&\multirow{2}{*}{D}&\multirow{2}{*}{$\times$8}&\multirow{2}{*}{48--20} &\SI{32.6}{\percent}&1.89\\
        B1 & & & & & &\SI{31.6}{\percent}&1.91\\
        \midrule
        C0 & \multirow{3}{*}{SK Hynix~\cite{datasheetHMAA4GU6AJR8N}} & \multirow{3}{*}{4Gb}&\multirow{3}{*}{F}&\multirow{3}{*}{$\times$8}&\multirow{3}{*}{51--20} &{\SI{35.3}{\percent}}&{1.89}\\
        C1 & & & & & &\SI{38.4}{\percent}&1.88\\
        C2 & & & & & &\SI{36.1}{\percent}&1.96\\
        \bottomrule
    \end{tabular}
    \begin{flushleft}
    $^*$ {HiRA Cov. stands for HiRA coverage results, presented in \secref{sec:experiments_coverage}.}\\
    $^{**}$ {Norm. $N_{RH}$ is {the} normalized RowHammer threshold, shown in \secref{sec:experiments_chargerestoration}.}\\
    \agyarb{Table~\ref{tab:detailed_info} in Appendix A shows the minimum and the maximum values for both HiRA Cov. and Norm. $N_{RH}$ across all tested rows per DRAM module.}
    \end{flushleft}
    \label{tab:dram_chip_list}
\end{table}

We conduct experiments {in three steps (\secref{sec:experiments_coverage}-\secref{sec:variation_across_banks})} to evaluate the {feasibility,} reliability,  {benefits} and limitations of HiRA on real DRAM chips.}

\subsection{Testing Infrastructure}
\begin{figure}[!ht]
    \centering
    \includegraphics[width=\linewidth]{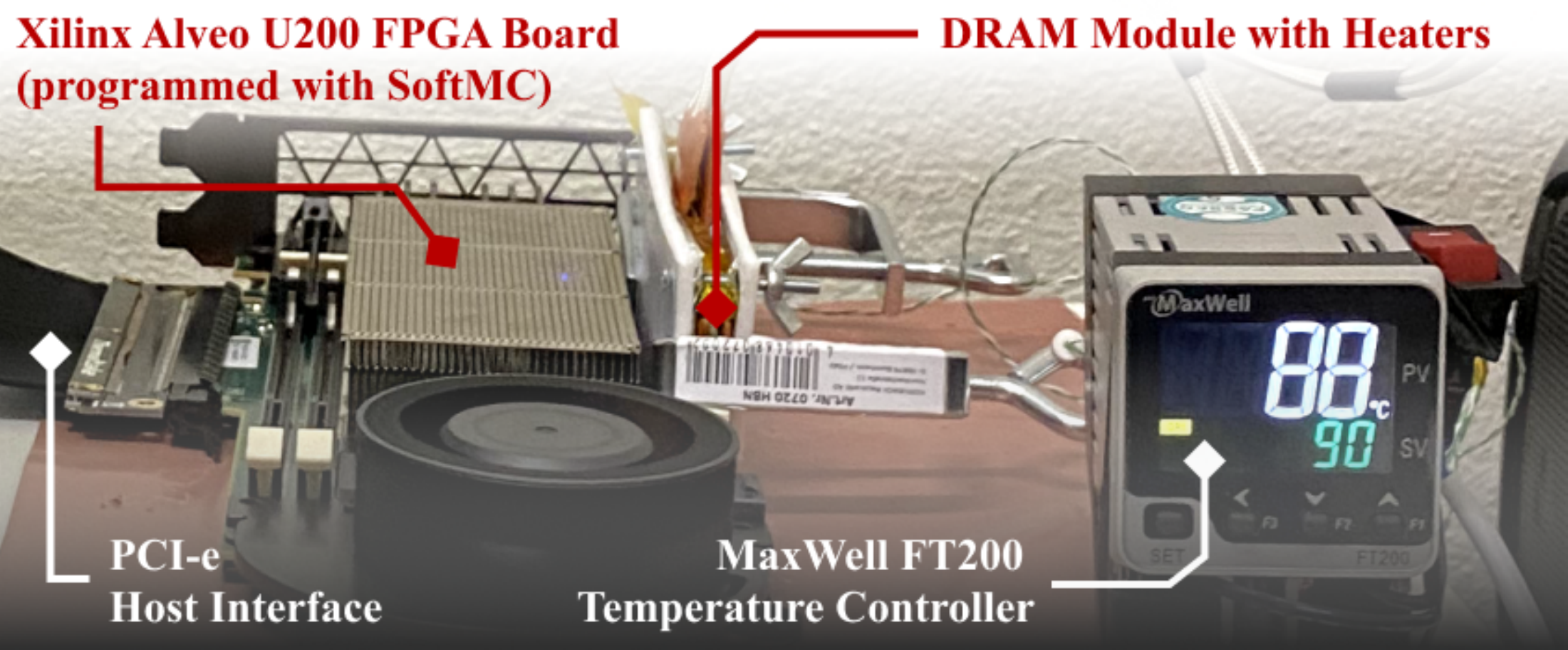}
    \caption{{SoftMC Infrastructure: Xilinx Alveo U200 FPGA board~\cite{alveo}, programmed with a DDR4 version of SoftMC~\cite{softmcgithub, hassan2017softmc}, PCI-e host interface, DRAM module clamped with heater pads, and MaxWell FT200 temperature controller~\cite{maxwellFT200}}}
    \label{fig:softmc}
\end{figure}

We conduct experiments {on \param{\chipcnt{}} real DRAM chips\footnote{\label{fn:testedrows}Due to time limitations, we conduct our tests on the 1)~first 2K, 2)~last 2K, and 3)~middle 2K rows of {Bank 0} in each DRAM chip, similar to~\cite{kim2014flipping, orosa2021deeper, yaglikci2022understanding}.}} using a modified version of SoftMC~\cite{hassan2017softmc, softmcgithub} {that can support DDR4 modules}. \figref{fig:softmc} shows a picture of our experimental setup.
We {{use the}} Xilinx Alveo {U200 FPGA} board~{\cite{alveo},} {programmed} {with} SoftMC {to} precisely issue {DRAM} command{s}.{\footnote{{SoftMC works with a minimum clock cycle of \SI{3}{\nano\second} on Alveo U200~\cite{alveo} and thus issues a DRAM command every \SI{1.5}{\nano\second} in {the} double data rate domain.}}} {The host machine generates the sequence of DRAM commands that we issue to the DRAM {module}.} To avoid {fluctuations} in ambient temperature, we {place the DRAM module clamped with a pair of heaters on both sides. The heaters are controlled by a MaxWell FT200~\cite{maxwellFT200}} temperature controller that {keeps} DRAM chips at $\pm$\param{\SI{0.1}{\celsius}} neighborhood of the {target} temperature.

\noindent
\textbf{Data Patterns.} {Our tests} use four data patterns {that are used by prior works~\cite{khan2014efficacy,liu2013experimental,patel2017reaper, kim2020revisiting, chang2016understanding, chang2017understanding, lee2017design,khan2016parbor,khan2016case,khan2017detecting,lee2015adaptive}}: 1)~all ones {($0xFF$)}, 2)~all zeros ($0x00$), 3)~alternating ones and zeros, i.e., checkerboard ($0xAA$), and 4)~the inverse checkerboard ($0x55$).

\noindent
\textbf{Disabling Sources of Interference.}
To directly observe whether HiRA reliably works {at the} circuit-level,
we {disable} all {known} sources of interference {(i.e., we prevent other DRAM error mechanisms (e.g., retention errors~\cite{liu2013experimental,khan2014efficacy,meza2015revisiting,patel2017reaper,qureshi2015avatar}) or error correction from interfering with a \hiraop{}'s results)} {in three steps, similar to prior works~\cite{orosa2021deeper, kim2020revisiting, yaglikci2022understanding}.} 
First, we disable all DRAM self-regulation events {(e.g., DRAM Refresh{)}} {and error mitigation mechanisms (e.g., error correction codes and RowHammer defense mechanisms)}~\cite{jedec2017ddr4,hassan2017softmc,ultrascale}) except {calibration related events (e.g., ZQ calibration, {which is required}} for signal integrity~\cite{jedec2017ddr4,hassan2017softmc}). {Second, we {conduct each} test within a relatively short period of time {(\SI{10}{\milli\second})} such that we do \emph{not} observe retention errors}. 
{Third, we conduct each test for ten iterations to reduce noise in our measurements.}

\subsection{HiRA's Coverage}
\label{sec:experiments_coverage}
{\gls{hira} works if the two rows that \gls{hira} opens do \emph{not} corrupt each other's data. Therefore, it is important to carefully choose two DRAM rows for \gls{hira} such that the rows are electrically isolated from each other, i.e., do \emph{not} share a bitline or sense amplifier. {The \emph{goal} of our first experiment is to find all combinations of DRAM row pairs that \gls{hira} can {concurrently} activate.} 
{To this end, we define HiRA's \emph{coverage} for a given row ($RowA$) {in a given bank {($BankX$)}} as the {fraction} of other DRAM rows {within $BankX$} which \gls{hira} can {reliably} activate {concurrently with} $RowA$.}
Algorithm~\ref{algorithm:hira} shows the experiment} {to find HiRA's \emph{coverage {for $RowA$.}}
{{To test a pair of DRAM rows $RowA$ and $RowB$ within $BankX$, first, we initialize the two rows using inverse data patterns (lines 7{--}8). Second, we perform \gls{hira} (lines 11--13) and close both rows (line 16). Third, we check whether there is a bit flip in either of the rows (lines 19{--}20). Fourth, if} performing \gls{hira} does \emph{not} cause bit flips in {either} of the rows for any tested data pattern, we increment a counter called $row\_count$ {(line 25). Fifth, we calculate} \gls{hira}'s coverage for $RowA$ {as the fraction of $RowB$s that \gls{hira} can concurrently activate with $RowA$ (line 26).}}
}

\SetAlFnt{\footnotesize}
\RestyleAlgo{ruled}
\begin{algorithm}
\caption{Testing HiRA's Coverage {for {a given} RowA}}\label{algorithm:hira}
  \DontPrintSemicolon
  \SetKwProg{Fn}{Function}{:}{}
  
  \For{RowA in Tested Rows {in BankX}}{
    {row\_count = 0}\;
    \For{RowB in Tested Rows {in BankX}}{
        success = True\;
        \For{datapattern in [$0xFF$, $0x00$, $0xAA$, $0x55$]}
        {
        \textbf{\emph{\# Initialize {the two rows with} inverse data patterns}}\;
        initialize($RowA$, datapattern)\;
        initialize($RowB$, !datapattern)\;
        \;
        \textbf{\emph{\# Perform HiRA}}\;
        act({$BankX$,}$RowA$, wait=$t_{1}$)\;
                pre({$BankX$,} wait=$t_{2}$)\;
        act({$BankX$,}$RowB$, wait=$t_{RAS}$)\; 
                \;
        \textbf{\emph{\# Clos{e} both rows}}\;
        pre({$BankX$,} wait=$t_{RP}$)\;
        \;
        \textbf{\emph{\# Read back the two rows and check for bit flips}}\;
        $RowA$\_pass = compare\_data(datapattern, $RowA$)\; 
        $RowB$\_pass = compare\_data(!datapattern, $RowB$)\;
        \;
        \textbf{\emph{\# Fail if there is at least one bit flip}}\;
        \lIf{!($RowA$\_pass AND $RowB$\_pass)}{success = false}
        }
        \lIf{success == true}{{row\_count}++}
    }
    {HiRA\_coverage[$RowA$] = row\_count / NumberOfTestedRows}\;
  }
\end{algorithm}

\begin{figure}[!t]
    \centering
    \includegraphics[width=\linewidth]{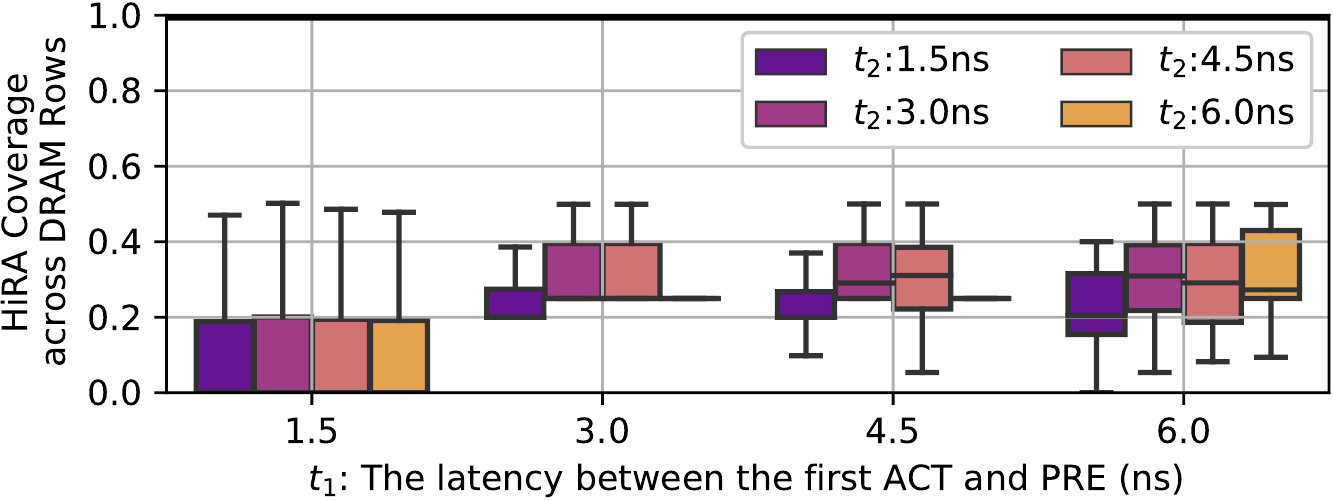}
    \caption{\gls{hira}'s coverage {across tested DRAM rows} for different $t_1$ (x-axis) and $t_2$ ({colored boxes}) timing parameter {combinations}}
    \label{fig:coverage}
\end{figure}

{\figref{fig:coverage} shows
the distribution of {\gls{hira} coverage across} {tested DRAM rows{\footref{fn:testedrows}}}
in a box and whiskers plot\footnote{\label{fn:boxplot}{A box-and-whiskers plot emphasizes the important metrics of a dataset’s distribution. The box is lower-bounded by the first quartile (i.e., the median of the first half of the ordered set of data points) and upper-bounded by the third quartile (i.e., the median of the second half of the ordered set of data points).
The \gls{iqr} is the distance between the first and third quartiles (i.e., box size).
Whiskers show the minimum and maximum values.}}  {for different combinations of $t_1$ {(x-axis)} and $t_2$ {({colored boxes})} timing parameters}.
{The y-axis shows \gls{hira} coverage across tested rows.}
}

We make three observations from \figref{fig:coverage}.
{First, if $t_1$ is \SI{3}{\nano\second} or \SI{4.5}{\nano\second}, a {given} DRAM row's refresh operation can {always} be performed concurrently with at least another DRAM row's {refresh or activation} (i.e., there are no DRAM rows with a \gls{hira} coverage of $0$) {for all tested $t_2$ values.}}
{Second,
{\gls{hira} reliably parallelizes a tested row's {refresh operation} {with refresh or {activation of} {any of the} \covref{}}}
of the other rows\footnote{{The minimum \gls{hira} coverage we observe across all tested rows is \SI{25}{\percent} when $t_1$ is \SI{3}{\nano\second} and $t_2$ is either \SI{3}{\nano\second} or \SI{4.5}{\nano\second}.}} when $t_1$ is \SI{3}{\nano\second} and $t_2$ is either \SI{3}{\nano\second} or \SI{4.5}{\nano\second}.
{Third, we observe that {\gls{hira} coverage} can be $0$ if $t_1$ is chosen too small (e.g., \SI{1.5}{\nano\second}) or too large (e.g., \SI{6}{\nano\second}), {meaning that at least one tested DRAM row's refresh \emph{cannot} be concurrently performed with refresh{ing or activating} another tested DRAM row.} We hypothesize that}
this happens because {1)~\SI{1.5}{\nano\second} is \emph{not} long enough to enable sense amplifiers and 2)~\SI{6}{\nano\second} is \emph{not} short enough for $t_1$ to interrupt row activation before the local row buffer is connected to the bank I/O or \SI{1.5}{\nano\second} is too short for $t_2$ to disconnect $RowA$'s local row buffer from the bank I/O.} Both design-{induced variation}~\cite{lee2017design} and manufacturing process-induced variation~\cite{chang2016understanding, lee2017design} in row activation latency can cause this behavior.
From these {three} observations, we conclude that it is possible to refresh {a given DRAM row concurrently} with {refreshing or activating} \SI{32}{\percent} of the other DRAM rows on average when both $t_1$ and $t_2$ are \SI{3}{\nano\second}.} 
{When \gls{hira} is used with the configuration of $t_1 = t_2 = 3ns$, the latency of refreshing two rows is \emph{only} \SI{38}{\nano\second} ($t_1 + t_2 + t_{RAS}$), while {refreshing two rows with standard DRAM commands takes} \SI{78.25}{\nano\second} for 1)~restoring the charge of the first row ($t_{RAS}$), 2)~precharging bitlines to prepare for second activation ($t_{RP}$), and 3)~restoring the charge of the second row ($t_{RAS}$).{\footref{fn:reftworows}}}
{Therefore, \gls{hira} reduces the latency of refreshing two rows by \SI{51.4}{\percent}.}

\subsection{{Verifying} HiRA's {Second Row Activation}}
\label{sec:experiments_chargerestoration}
Observing \emph{no bit flips} for a pair of rows tested using Algorithm~\ref{algorithm:hira} indicates either of {the} two cases: 1)~\gls{hira} successfully works or 2)~\gls{hira} activates only {the first} row because {the} DRAM chip simply ignores the second \gls{act} command. {The \emph{goal} of our second experiment is to} test whether the DRAM chip ignores or performs \gls{hira}'s second row activation command.
{To this end,} we hammer the two adjacent rows\footnote{{DRAM manufacturers {use {DRAM-internal} mapping} {schemes} to internally translate memory-controller-visible row addresses to physical row {addresses~\cite{kim2014flipping, smith1981laser, horiguchi1997redundancy, keeth2001dram, itoh2013vlsi, liu2013experimental,seshadri2015gather, khan2016parbor, khan2017detecting, lee2017design, tatar2018defeating, barenghi2018software, cojocar2020rowhammer,  patel2020beer, yaglikci2021blockhammer, orosa2021deeper}{,} which can vary across different DRAM modules}. We reconstruct this mapping using single-sided RowHammer, similar to prior works~\cite{kim2014flipping, kim2020revisiting, orosa2021deeper, yaglikci2022understanding}, so {that} we can hammer aggressor rows that are physically adjacent to a victim row.}} (i.e., aggressor rows) of a given victim row to induce RowHammer bit flips in the victim row (i.e., double-sided RowHammer~\cite{kim2014flipping, kim2020revisiting}). {During the test}, we try refreshing the victim row by using {\gls{hira}'s} \emph{second} row activation command. We hypothesize that {if \gls{hira}'s second row activation is \emph{not} ignored {(i.e., if \gls{hira} works)}, then}
{the minimum number of aggressor row activations required to induce the first RowHammer bit flip, i.e., RowHammer threshold {(\secref{sec:background_rowhammer})}, increases compared to the RowHammer threshold measured \emph{without} using \gls{hira}.}
{W}e \emph{measure} RowHammer threshold {of a given victim row} via binary-search (similar to prior works~\cite{kim2020revisiting, orosa2021deeper, yaglikci2022understanding}).
{Algorithm~\ref{algorithm:rowhammer} shows how we perform a RowHammer {test} {for a given victim row} at a given hammer count ($HC$) with and without a \gls{hira} operation.}

\SetAlFnt{\footnotesize}
\RestyleAlgo{ruled}
\begin{algorithm}[!ht]
\caption{{{Verifying} \gls{hira}'s Second {Row Activation}}}\label{algorithm:rowhammer}
  \DontPrintSemicolon
  \SetKwProg{Fn}{Function}{:}{}
  
  \For{with\_HiRA in [False, True]}{
    \textbf{\emph{{\# Step 1: Initialize DRAM rows}}}\;
    \textbf{\emph{\# Initializ{e} the victim row with the specified data pattern}}\;
    initialize(victim\_row, datapattern)\;
    \textbf{\emph{\# Initializ{e a} dummy row for \gls{hira}{'s first ACT}}}\;
    initialize(\gls{hira}\_dummy\_row, !datapattern)\;
    \textbf{\emph{\# Initializ{e} the two aggressor rows with {inverse} data pattern}}\;
    initialize(aggr\_row\_a, !datapattern)\;
    initialize(aggr\_row\_b, !datapattern)\;
        \;
    \textbf{\emph{\# Step {2}: Hammer each aggressor row $HC/2$ times}}\;
    \For{for act\_cnt = 0 to $HC/2$}{
        act({BankX,} aggr\_row\_a, wait=$t_{RAS}$)\; 
        pre({BankX,} wait=$t_{RP}$)\;
        act({BankX,} aggr\_row\_b, wait=$t_{RAS}$)\;
        pre({BankX,} wait=$t_{RP}$)\;
    }
    \textbf{\emph{\# Step {3}: Perform HiRA or wait}}\;
    \If{with\_HiRA}{
        act({BankX,} \gls{hira}\_dummy\_row, wait=$t_{1}$)\;   
        pre({BankX,} wait=$t_{2}$)\;
        act({BankX,} victim\_row, wait=$t_{RAS}$)\; 
        pre({BankX,} wait=$t_{RP}$)\;
    }\Else{ \textbf{\emph{{\# Without HiRA}}}\;
        wait($t_{1}$+$t_{2}$+$t_{RAS}$+$t_{RP}$);\;
    }
    \textbf{\emph{\# Step {4}: Hammer each aggressor row $HC/2$ times}}\;
    \For{for act\_cnt = 0 to $HC/2$}{
        act({BankX,} aggr\_row\_a, wait=$t_{RAS}$)\; 
        pre({BankX,} wait=$t_{RP}$)\;
        act({BankX,} aggr\_row\_b, wait=$t_{RAS}$)\;
        pre({BankX,} wait=$t_{RP}$)\;
    }
    {\textbf{\emph{\# Step {5}: Check for bit flips on the victim row}}\;}
    {bitflips} = {check\_bitflips(datapattern, victim\_row)}\;
  }
\end{algorithm}

{We conduct the RowHammer test in {{five}} steps.}
{First, we initialize four DRAM rows in a given DRAM bank ({BankX}): the given victim row, a dummy row which \gls{hira} can concurrently refresh with the given victim row, and the two aggressor rows. We initialize the victim row with the specified data pattern and the other three rows with the inverse data pattern (lines {3--9}).}
{Second, we hammer {each aggressor row $HC/2$} times {({lines 12--16})}. Third, we either perform {{(lines 19--23)}} a \gls{hira} operation {({\emph{with}} \gls{hira})} or {wait {(line {26})} exactly the same amount of time as performing \gls{hira} would take {({\emph{without}} \gls{hira})}}. Fourth, we hammer {both aggressor} rows $HC/2$ times {(lines {30--33})}.} {If \gls{hira}'s second row activation is \emph{not} ignored,} {then the victim row would be refreshed, and thus} we {would} observe a significant increase in measured RowHammer threshold values {in the test {\emph{with} HiRA}, compared to the {test \emph{without} HiRA}}. {Fifth, we read the victim row to check if the RowHammer test causes any bit flip (line 36).}

\begin{figure}[!ht]
    \vspace{1em}
    \centering
    \includegraphics[width=\linewidth]{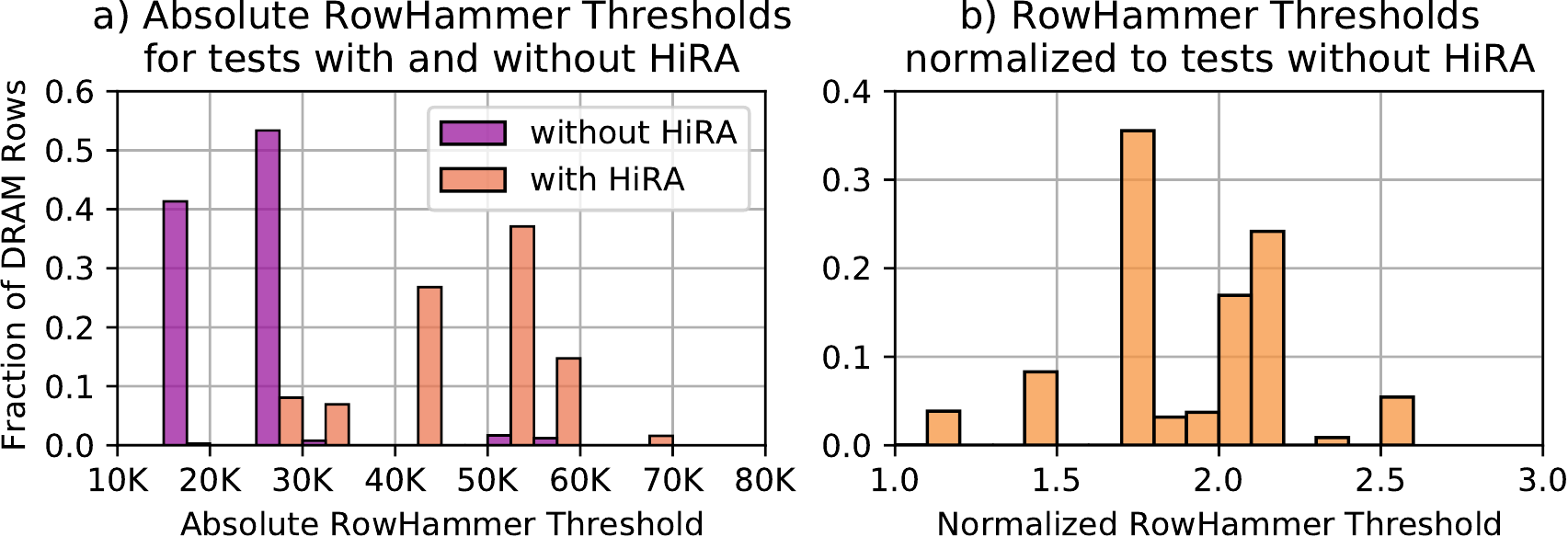}
    \caption{Variation in RowHammer threshold {due to} \gls{hira}{'s second {row} activation}}
    \label{fig:norm_hcfirst}
\end{figure}

\pagebreak
\figref{fig:norm_hcfirst} {shows} how {a DRAM row's} RowHammer threshold varies when the row is refreshed using \gls{hira}. \figref{fig:norm_hcfirst}{a and~\ref{fig:norm_hcfirst}b show the histogram of absolute and normalized RowHammer threshold values, respectively.} 
We report the normalized values relative to {tests without} \gls{hira}. 

{We make two observations. First, \figref{fig:norm_hcfirst}a shows that RowHammer threshold is $27.2K$ / $51.0K$ on average across tested rows when tested \emph{without} / \emph{with} \gls{hira}.
Second, \figref{fig:norm_hcfirst}b shows that} RowHammer threshold increases {by $1.9\times$ on average across tested DRAM rows and} by more than 1.7$\times$ for the vast majority {(\SI{88.1}{\percent})} of {tested} rows. {Based on these {two observations},} we conclude that {\gls{hira} works in 56 tested DRAM chips (\tabref{tab:dram_chip_list}) such that} \gls{hira}{'s second row activation, targeting the victim row, is \emph{not} ignored, and thus the victim row is successfully activated} {concurrently} with {the dummy row}.

\subsection{Variation Across DRAM Banks}
\label{sec:variation_across_banks}
To {investigate} the variation in \gls{hira}'s {coverage and verify \gls{hira}'s second row activation} across DRAM banks, {we repeat {the} tests {that we explain in \secref{sec:experiments_coverage} and \secref{sec:experiments_chargerestoration}} for {\emph{all}} 16 banks {of}} three DRAM modules{: A0, B0, and C0 {(Table~\ref{tab:dram_chip_list})}.}

\subsubsection{HiRA's Coverage}
We observe that the pairs {of rows that {\gls{hira}} can {concurrently refresh and activate}} are \emph{identical} across {all 16} DRAM banks in all three modules.
{Based on this observation, we {hypothesize} that \gls{hira}'s coverage {largely} depends on the DRAM circuit design, which should be {a design-induced phenomenon} across all DRAM banks.}

\subsubsection{{Verifying} HiRA's {Second Row Activation}}
\begin{figure}[!ht]
    \centering
    \includegraphics[width=\linewidth]{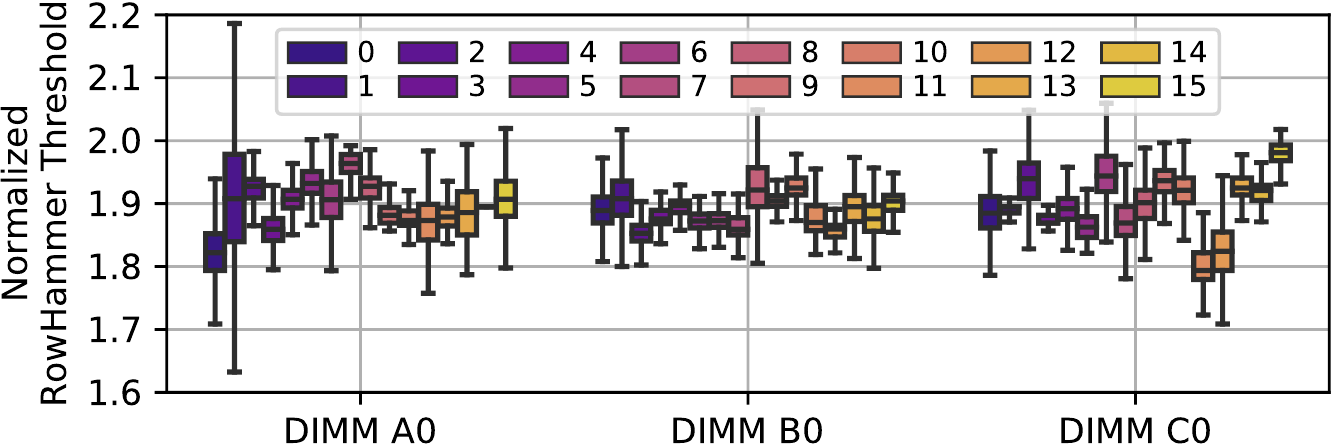}
    \caption{Variation in {normalized RowHammer threshold across banks {in three modules} {due to HiRA's second row activation}}}
    \label{fig:bank_variation}
\end{figure}

{To verify that HiRA's second row activation works across all 16 DRAM banks, we repeat the tests shown in \algref{algorithm:rowhammer}.
\figref{fig:bank_variation} shows how a DRAM row's RowHammer threshold varies when the victim row is activated using \gls{hira}'s second activation during a RowHammer attack (similar to \figref{fig:norm_hcfirst}b). The x-axis and different box colors show the module's name and DRAM bank, respectively. The y-axis shows the measured RowHammer threshold in the tests \emph{with \gls{hira}}, normalized to the tests \emph{without \gls{hira}}.}
{E}ach box in \figref{fig:bank_variation} shows the distribution's \gls{iqr} and whiskers show the minimum and maximum values.{\footref{fn:boxplot}}

{We make three observations from \figref{fig:bank_variation}. First, the {normalized RowHammer threshold values are larger than \SI{1.56}{\times} {across \emph{all} banks in \emph{all} three DRAM modules}. Second, RowHammer threshold increases by \SI{1.89}{\times}, average{d} across all banks in all three modules{,} when the victim row is refreshed using \gls{hira}. {Third,} the average RowHammer threshold increase in a DRAM bank varies between \SI{1.80}{\times} and \SI{1.97}{\times} {across \emph{all} banks in \emph{all} three modules}.
{Therefore, we conclude that \gls{hira}'s second row activation is \emph{not} ignored in \emph{any} bank.}}}

% \pagebreak
\setstretch{0.95}
\section{HiRA-MC: HiRA Memory Controller}
\label{sec:implementation}

{\Gls{hirasched} {aims} to improve overall system performance. {To do so, \gls{hirasched} queues each periodic and preventive refresh request with a deadline and takes one of three possible actions in decreasing priority order: 1)~{concurrently perform a} refresh operation with a \memacc{} {(refresh-access parallelization)} before the refresh operation's deadline; 2)~{concurrently perform a} refresh operation with another refresh operation {(refresh-refresh parallelization)} if no \memacc{} can be parallelized until the refresh operation's deadline; or 3)~perform a refresh operation {by} its deadline if the refresh {operation \emph{cannot}} be {concurrently performed} with {a} \memacc{} {or} a{nother} refresh.} \gls{hirasched} intelligently schedules refresh operations from within the memory controller \emph{without} {requiring any} modification to off-the-shelf DRAM chips.}

\begin{figure}[t]
    \centering
        \includegraphics[width=1\linewidth]{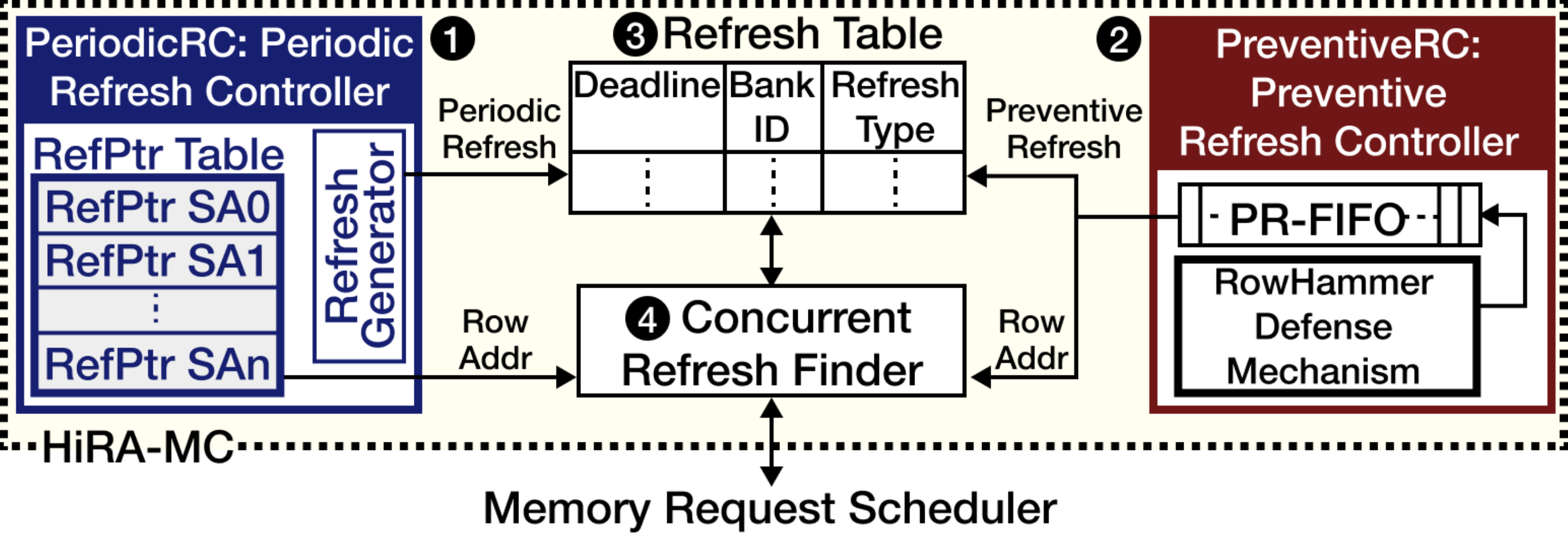}
    \caption{\gls{hirasched}'s components}
    \label{fig:mechanism}
\end{figure}

\figref{fig:mechanism} shows \gls{hirasched}'s components and {their} interaction with the memory request scheduler. {First, we give an overview of \gls{hirasched} where we introduce its components.} Then, we explain how \gls{hirasched}'s {components} interact with the memory request scheduler in {performing} four {key operations}.

\noindent
{\textbf{\gls{hirasched} Overview.}} {\gls{hirasched} consists of four main components: \emph{Periodic Refresh Controller ({PeriodicRC})}, \emph{Preventive Refresh Controller ({PreventiveRC})}, \emph{Refresh Table}, and \emph{{Concurrent} Refresh Finder}.} 
{\circled{1}~\emph{{PeriodicRC}} generates {a \emph{periodic}} refresh request {for each DRAM row} to maintain data integrity {in {the} presence of} DRAM cell charge leakage. {To leverage \gls{hira}'s {subarray-level} parallelism, {{PeriodicRC}} maintains a table called \emph{RefPtr table}. RefPtr table has an entry per subarray, which contains a pointer to the next row to be refreshed within the corresponding subarray.} 
\circled{2}~\emph{{PreventiveRC}} employs a refresh-based RowHammer defense mechanism {(e.g., PARA~\cite{kim2014flipping})} to generate {a} \emph{preventive} refresh request {for a {victim} DRAM row.}
{{There {might not be \emph{any} memory access} requests {that can be} parallelize{d} with a periodic or preventive refresh} {when} {the} refresh request is generated {(i.e., there {might not be \emph{any}} load or store memory requests waiting to be served by the memory controller)}. To address this issue,} both {PeriodicRC} and {PreventiveRC} {assign {each refresh} request a \emph{deadline} that specifies the {timestamp until which} the refresh request \emph{must} be {performed.
The deadline is determined using a configuration parameter called \gls{trrslack}.}
{\circled{3}~\emph{The Refresh Table} stores generated periodic and preventive refresh requests along with their deadline, {target} bank id, and {refresh type} (invalid, periodic, or preventive).}}}
{\circled{4}~\emph{The {Concurrent} Refresh Finder} identifies the refresh requests that can be parallelized with \memacc{} requests among the refresh requests stored in the Refresh Table.} 
To serve a refresh {request} {concurrently} with a \memacc{} request, {the {Concurrent} Refresh Finder} observes the \memacc{} requests that the memory request scheduler\footnote{{The \emph{memory request scheduler} is the component {that is} responsible for scheduling {DRAM requests, using a scheduling algorithm (e.g., FR-FCFS~\cite{rixner00, zuravleff1997controller} or PAR-BS~\cite{mutlu2008parbs})}, {and issuing} DRAM commands to serve {those} requests{.} 
}} issues. If there is a refresh request {that} can be parallelized with {a} {memory} request, the {Concurrent} Refresh Finder replaces the {memory} request's row activation command with a \gls{hira} operation, such that \gls{hira}'s first ACT targets the {row {that needs to be refreshed} and \gls{hira}'s second ACT targets the row {that needs to be accessed}}. 
If \gls{hirasched} \emph{cannot} perform a pending refresh request {concurrently} with a \memacc{} until the refresh request's deadline, {the {Concurrent} Refresh Finder} searches for another refresh request {within the Refresh Table} to parallelize {the refresh request with}. If possible, \gls{hirasched} performs a \gls{hira} operation to {concurrently} refresh two rows. {If {the} refresh {request} \emph{cannot} be parallelized with an access or another refresh}, \gls{hirasched} activates the row that needs to be refreshed and precharges {the bank} using nominal timing parameters.  

\subsection{{HiRA-MC: Key Operations}}

\subsubsection{Generating Periodic Refresh Requests} 
{The Periodic Refresh Controller} periodically {{generates refresh requests}}. {{{PeriodicRC}} faces two main challenges in scheduling \gls{hira} operations due to two fundamental differences between \gls{hira} and \gls{ref} operations. 
First, a \gls{ref} command refreshes several rows in a DRAM bank as a batch~\cite{jedec2017ddr4,jedec2020ddr5,liu2012raidr}{. In contrast,} using the \gls{hira} operation, the memory controller needs to issue an $ACT$ command for each {refreshed} DRAM row. Therefore, using \gls{hira} increases {DRAM} bus utilization compared to using \gls{ref} commands. For example, to refresh 64K rows in a bank of a DDR4 DRAM chip in \SI{64}{\milli\second}, the memory controller issues 8K \gls{ref} commands (once every \SI{7.8}{\micro\second}~\cite{jedec2017ddr4}), {indicating} that each \gls{ref} command refreshes eight rows {in one bank}. To ensure the same refresh rate {as the baseline}, {{PeriodicRC}} schedules 64K \gls{hira} operations (once every \SI{975}{\nano\second}).} 
{Second, issuing a \gls{ref} command triggers refresh operations in \emph{all} banks in a rank {(assuming all-bank refresh, as in DDR4~\cite{jedec2017ddr4})}. {In contrast}, \gls{hira} operations are performed separately for each DRAM bank because they use already defined $ACT$ and $PRE$ commands {at} row- and bank-level, respectively.} 
Therefore, {frequently issued} \gls{hira} command sequence{s} can occupy the command bus more than \gls{ref} commands. {{{For example, \gls{hirasched} needs} to perform 128 \gls{hira} operations {to refresh the same number of rows as one \gls{ref} command does {in current systems},} assuming {that a single \gls{ref} command refreshes eight rows from each of the 16 banks in a rank as in DDR4~\cite{jedec2017ddr4}.}}} 
To avoid overwhelming the command bus with \gls{hira} {operations}, {{PeriodicRC}} spreads the command bus pressure of \gls{hira} command sequences over time {by generating} {\gls{ref} {requests}} for each bank {with the same period, starting at different time offsets{. For example, assuming that
1)~each bank receives a \gls{ref} request once in every \SI{975}{\nano\second} and 2)~there are 16 banks,}
{{PeriodicRC}} {generates a refresh request every \SI{60.9}{\nano\second} (\SI{975}{\nano\second}/16 banks) targeting a different bank}.}
{{PeriodicRC}} inserts {the generated} {\gls{ref} request} into the Refresh Table \atb{with {the request's}} {1)~\emph{deadline}, which is a timestamp pointing {to} the time that is \gls{trrslack} later than the request's generation time,}
{2)~\emph{BankID}, which is the target bank of the refresh request, and 3)~{\emph{refresh type}}, which is set to \emph{Periodic} to indicate that the refresh request will perform a \emph{periodic} refresh operation.}

\subsubsection{Generating {Preventive} Refresh Requests {for {RowHammer}}}
\gls{hirasched} is not a RowHammer defense mechanism by itself, but {it provides parallelism support for} all memory controller-based {preventive} refresh mechanisms, which mitigate {the} RowHammer {effect} on victim rows by {timely} refreshing the victim rows~\rhdefrefresh{}.
{\gls{hirasched} overlaps} the latency of a {preventive} refresh operation with {another periodic/preventive refresh} or a \memacc{}. 
{To {do so, {{PreventiveRC}} generates preventive refreshes with a large enough} \gls{trrslack} without compromising the security guarantees {of RowHammer defense mechanisms}. To achieve this,}
{{PreventiveRC} assumes the worst case{,} where an attack fully exploit{s} \gls{trrslack} to maximize {the} {hammer} count such that the attack performs \gls{trrslack}/\gls{trc} additional activations during \gls{trrslack} (after the preventive refresh is generated and before it is performed). To account for {such case},}
{{{PreventiveRC}} {employs state-of-the-art RowHammer defense mechanisms\cite{park2020graphene, kang2020cattwo, lee2019twice, seyedzadeh2018cbt, qureshi2022hydra, kim2022mithril,kim2014flipping, son2017making, you2019mrloc} {with a slightly increased aggressiveness in performing preventive refreshes}.
When implemented in {PreventiveRC}, the \emph{counter-based} RowHammer defense mechanisms~\cite{park2020graphene, kang2020cattwo, lee2019twice, seyedzadeh2018cbt, qureshi2022hydra, kim2022mithril} should be configured such that the mechanism triggers a preventive refresh at a {hammer} count that is \gls{trrslack}/\gls{trc} activations smaller than the mechanism's original {hammer} count threshold {(typically, the {RowHammer threshold} of the DRAM module)}. Therefore, even if the attack performs the maximum number of activations after the preventive refresh is generated, the total {hammer} count does not exceed the RowHammer defense mechanism's original threshold before the preventive refresh is performed.
When implemented in {PreventiveRC}, the \emph{probabilistic} RowHammer defense mechanisms~\cite{kim2014flipping, son2017making, you2019mrloc} should be configured with an increased probability threshold to maintain the same security guarantees as the original {mechanism} in the presence of \gls{trrslack}. \secref{sec:security_analysis} explains how to increase the probability threshold to account for \gls{trrslack}.}}

When the employed RowHammer defense mechanism generates a {preventive} refresh {request}, {{PreventiveRC}} 1)~enqueues the refresh operation in a first-in-first-out queue, called \emph{PR-FIFO} {(\circled{2} in \figref{fig:mechanism})}, 2)~creates an entry in the Refresh Table {(\circled{3} in \figref{fig:mechanism})} with the preventive refresh's deadline and bank id, and 3)~{sets the request's {refresh type} to \emph{Preventive} to indicate that the refresh request will perform a \emph{preventive} refresh operation}.

\subsubsection{Finding {Concurrent} Refresh Operations}
\label{sec:finding-concurrent-refresh}
\begin{figure}[ht]
    \centering
    \includegraphics[width=1\linewidth]{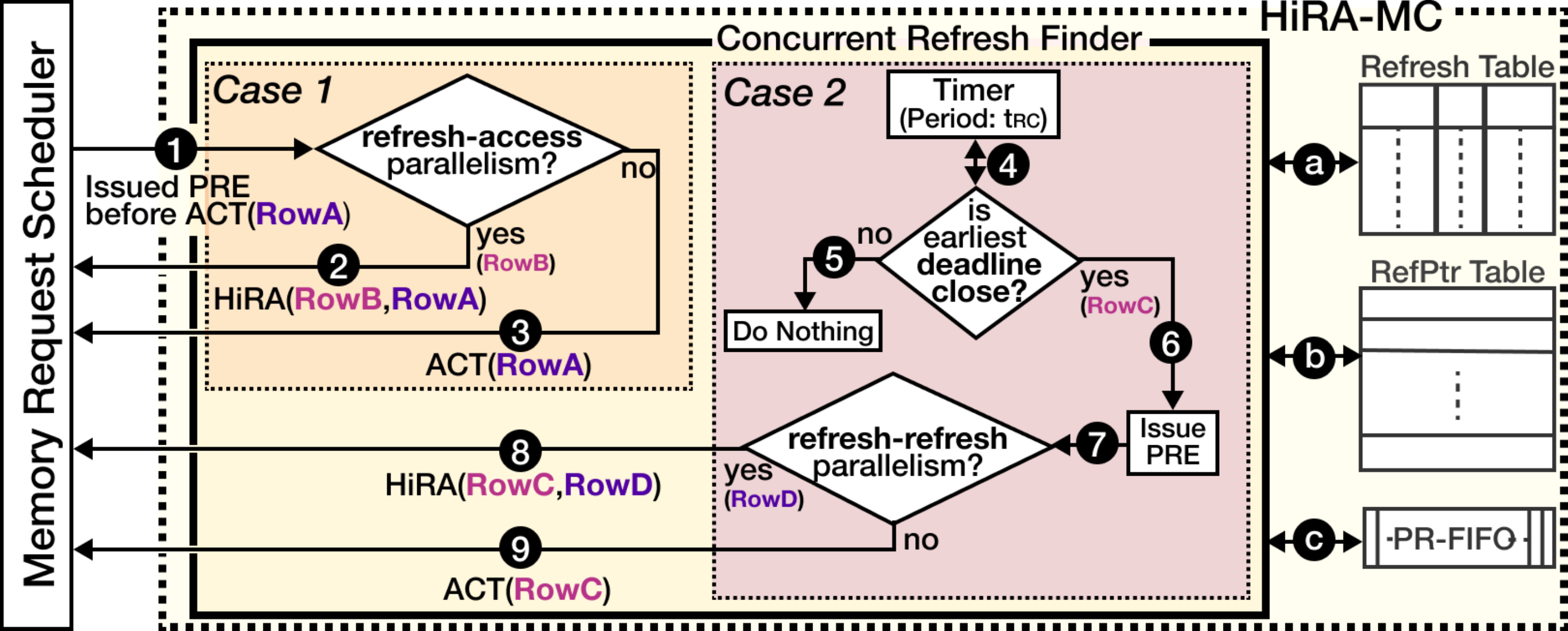}
    \caption{The Concurrent Refresh Finder's interaction with the memory request scheduler}
    \label{fig:mechanism_flowdiagram}
\end{figure}

{\figref{fig:mechanism_flowdiagram} shows how \gls{hirasched}'s {Concurrent} Refresh Finder interacts with the memory request scheduler {in two different cases: 1)~when the memory request scheduler issues a \gls{pre} command to prepare the bank for activating a $RowA$ (\circled{1} in \figref{fig:mechanism_flowdiagram}) and 2)~when an internal timer periodically initiates a process that performs refreshes by their deadline 
(\circled{4} in \figref{fig:mechanism_flowdiagram}).}}

\noindent
{\textbf{Case~1.}}
To find an opportunity {to} {concurrently refresh a DRAM row} with a \memacc{} {(refresh-access parallelism)}, {the {Concurrent} Refresh Finder} observes the commands that the memory request scheduler issues.
{When the memory request scheduler issues a \gls{pre} command to pre{charge} the DRAM bank {before} activat{ing} a DRAM row ($RowA$)~\circled{1}, the Concurrent Refresh Finder} {searches} the Refresh Table ({by iterating over the Refresh Table entries} in {the} order {of increasing deadlines) {to find a refresh operation} {that targets} the precharged bank}~(\circled{a}).

{{If the Refresh Table entry with the {earliest} deadline is a periodic refresh,}}
{{the Concurrent Refresh Finder} {accesses} the RefPtr Table~(\circled{b}) {to find} {a} subarray} 
{1)~{where} refreshing a DRAM row} can be parallelized with the activation {of $RowA$} {{and 2)~that has the smallest number of DRAM rows} refreshed within the current refresh window.}
By doing so, \gls{hirasched} aims {to} advanc{e} the refresh pointers of all subarrays in a balanced manner {while leveraging the subarray-level parallelism.}

{If the Refresh Table entry with the {earliest} deadline is a preventive refresh,}
{the Concurrent Refresh Finder} checks whether the request at the head of {P}R-FIFO can be {refreshed concurrently} with {activating $RowA$}~(\circled{c}).

If \gls{hirasched} finds a periodic or preventive refresh {{that targets} $RowB$} {and can be} {concurrently perform{ed}} with an {activation to $RowA$~(\circled{2})}, the memory request scheduler issues {a} \gls{hira} {operation} such that the first {and the second} \gls{act} {commands} target {$RowB$ and {$RowA$}, respectively{, so that $RowB$ is refreshed concurrently with activating $RowA$.}} 

If there is \emph{no} opportunity {{to}} {concurrently perform a periodic or preventive refresh with the activation of $RowA$ (\circled{3}),}
the memory request scheduler {issues} a regular \gls{act} {command} targeting $RowA$. {In this case, 1)~the {DRAM row} activation is performed \emph{without} leveraging \gls{hira}'s parallelism} 
{because refresh-access parallelism is not possible and 2)~the} {memory access request is prioritized over the queued refresh requests}
{to improve system performance when queued refresh requests can be delayed until their deadline.}

\noindent
{\textbf{Case 2.}}
{To guarantee that each periodic and preventive refresh is performed {timely (i.e., by its deadline)}, the Concurrent Refresh Finder periodically checks if there is a refresh operation that is close to its deadline (i.e., whose deadline is {earlier} than \gls{trc})~(\circled{4}). If there is {\emph{no} queued periodic or preventive refresh whose deadline is close~(\circled{5}), the Concurrent Refresh Finder does \emph{not} take any action{. In doing so, \gls{hirasched}} 1)~does \emph{not} interfere with the memory access requests and 
2)~{opportunistically leaves refresh requests in the Refresh Table such that {the refresh requests} can be concurrently performed with a memory access by their deadlines.}}

If there is a refresh {request {(targeting a $RowC$)}} that needs to be performed {within the next \gls{trc} time window~(\circled{6}),} the Concurrent Refresh Finder precharges {the target} bank {of the refresh operation} if {the bank} is open and~(\circled{7})~tries leveraging refresh-refresh parallelism {by searching for a queued refresh operation that can be concurrently performed with refreshing $RowC$.}
{If {there is a refresh request {(}targeting a $RowD${),} which can be concurrently performed with refreshing $RowC$ (refresh-refresh parallelism), \gls{hirasched} forces the memory request scheduler to issue a \gls{hira} operation such that the two activations of \gls{hira} target $RowC$ and $RowD$, {respectively}~(\circled{8}).}
{If such refresh-refresh parallelism is not possible~(\circled{9}),} the memory request scheduler issues a regular \gls{act} command {targeting $RowC$} to perform the refresh operation {because 1)~refresh-refresh parallelism is \emph{not} possible and 2)~delaying $RowC$'s refresh {would violate its deadline and could} have caused bit flips}.}}

\subsubsection{Maintaining the Parallelism Information}
{T}o know {if a} DRAM row can be {concurrently} activated with {another} DRAM row, the memory controller needs to {determine   
whether the two rows are located in subarrays {that} do \emph{not} share {bitlines} or sense {amplifiers}.}
The memory controller can {learn
which subarrays do not share {bitlines or a sense amplifiers} with another subarray {(i.e., determine the \emph{boundaries of a subarray})}} {in two ways. First, the memory controller can} perform a one-time reverse engineering (e.g., by testing for HiRA's coverage as we do in \secref{sec:experiments_coverage}). {Second}, DRAM manufacturers can {expose this information} to the memory controller by {using} mode status registers (MSRs)~\cite{jedec2017ddr4} in the DRAM chip.
{Once the memory controller obtains the subarray boundaries, it} maintains them
in a table {called \emph{{S}ubarray {P}airs {T}able (SPT)}} {implemented as an on-chip {SRAM} storage}. {{SPT has an entry for each subarray ($S_A$). {{This entry}} contains a list of subarrays which do \emph{not} share bitlines or sense amplifiers with $S_A$. Therefore, \gls{hira} operation can be performed targeting a DRAM row in $S_A$ and another DRAM row in any of the listed subarrays.}}

\subsection{Power Constraints}
Each refresh and row activation in a HiRA operation is counted as a row activation with respect to {the} \gls{tfaw} constraint {in DDRx DRAM chips}. We respect \gls{tfaw} in our performance evaluation, such that within a given \gls{tfaw}, at most four activations are performed in a DRAM rank {(as required by {DRAM} datasheets {(e.g.,~\cite{jedec2017ddr4,jedec2020ddr5,jedec2015lpddr4,jedec2020lpddr5})})}, {thereby ensuring} that the row activations are performed within the power budget of a DRAM rank.

\subsection{Compatibility with Off-the-{Shelf} DRAM Chips}
We experimentally demonstrate that \gls{hira} works on {all} 56 {real} DRAM chips {that we test} {(\secref{sec:characterization})};
and \gls{hirasched} does \emph{not} require any modifications to these {real} DRAM chips to {enable} refresh-refresh and refresh-access parallelization.

\subsection{Compatibility with {Different Computing} System{s}}
{We discuss \gls{hirasched}'s compatibility with three {types} of computing systems: {1) F}PGA-based systems {(e.g., PiDRAM \cite{olgun2021pidram})}, {2) c}ontemporary processors, and {3) s}ystems with programmable memory controllers~{\cite{tassadaq2014pmss, bojnordi2012pardis}}. First, {\gls{hirasched} can be easily integrated into all existing FPGA-based systems that use DRAM to store data~{\cite{olgun2021pidram, xilinx-data-center, xilinx-fpga}} {{by} implementing \gls{hirasched} in RTL}.}}
{Second, contemporary processors require modifications to their memory controller logic to implement \gls{hirasched}. Implementing \gls{hirasched} is a design-time decision that requires balancing manufacturing cost with periodic and {preventive} refresh overhead reduction benefits. 
{We show that \gls{hirasched} significantly improves system performance (\secref{sec:doduoref} and~\secref{sec:doduopara}) at low chip area cost (\secref{sec:hardware_complexity}) and thus {can be relatively easily integrated into contemporary processors}.}
Third, systems that employ programmable controllers~\cite{tassadaq2014pmss,bojnordi2012pardis} can be {relatively} easily modified to implement \gls{hirasched} {by} programming {the \gls{hira} operation and implementing \gls{hirasched}'s components using the ISA of programmable memory controllers~\cite{tassadaq2014pmss,bojnordi2012pardis}.}}

\section{Hardware Complexity}
\label{sec:hardware_complexity}

{We evaluate the hardware complexity of implementing \gls{hirasched} in a processor,} {using CACTI 7.0~\cite{cacti}} to model {\gls{hirasched}'s components (}Refresh Table{, RefPtr Table, and} {P}R-FIFO). We use {CACTI's} \SI{22}{\nano\meter} {technology node to model SRAM array{s} for each component's on-chip data storage.} {Table~\ref{table:scheduler-area} shows the area cost {and the access latency} of each component.}

\begin{table}[h] 
    \begin{threeparttable}
    \caption{{The} area {cost (per DRAM rank)} {and access latency of HiRA-MC's components}}
    \label{table:scheduler-area}
    \centering
    \footnotesize
    \begin{tabular}{l r r r}
        \textbf{\gls{hirasched} Component} & \textbf{Area ($mm^2$)} & \textbf{Area (\SI{}{\percent}{\textsuperscript{*}})} & \textbf{Access Latency}\\
    \hline
    Refresh Table & {$0.00031$} & <\SI{0.0001}{\percent} & \SI{0.07}{\nano\second}\\
    RefPtr Table & $0.00683$ & \SI{0.0017}{\percent} & \SI{0.12}{\nano\second}\\
    PR-FIFO & $0.00029$ & <\SI{0.0001}{\percent} & \SI{0.07}{\nano\second}\\
    {Subarray Pairs Table (SPT)} & {$0.00180$} & \SI{0.0005}{\percent} & \SI{0.09}{\nano\second}\\
    \hline
    {Overall} & {$0.00923$} & {\SI{0.0023}{\percent}} & {\textsuperscript{**}}{\SI{6.31}{\nano\second}}\\

    \end{tabular}
    \begin{tablenotes}
    \footnotesize
    \item {\textsuperscript{*}{Normalized to the die area of} a 22nm Intel processor~\cite{inteli7}}.
    \item \textsuperscript{**}Calculated as the {overall latency of serially accessing 1)~the SPT, 2)~the Refresh Table, {and 3)~the RefPtr Table for {68} times}} {(\secref{sec:hirasched-access-latency})}.
    \end{tablenotes}
    \end{threeparttable}
\end{table}

\noindent
\textbf{Refresh Table.} {In this analysis, we assume a \gls{trrslack} of 4\gls{trc} {because 1)~increasing \gls{trrslack} increases the hardware complexity of \gls{hirasched} (by increasing the number of entries in the Refresh Table and the PR-FIFO) and 2)~a \gls{trrslack} of 4\gls{trc} {already} provides {as large} performance benefit {as a \gls{trrslack} of 8\gls{trc}} (\secref{sec:doduoref} and~\secref{sec:doduopara}).}
Within a time window of {{4}}\gls{trc}, {\gls{hirasched} can generate at most} {{4}} {periodic {refresh requests per \emph{rank}} and {4} preventive} refresh requests per \emph{{b}ank} {(64 preventive refresh requests per \emph{rank})}. Therefore, a Refresh Table with {{68}} entries {per rank}  is enough to store all generated refresh requests.}
Each entry {consists} of 1) 1{0} bits to store the deadline,\footnote{
{A {10}-bit number can represent the number of clock cycles within a \gls{trrslack} of 4\gls{trc} (\SI{185}{\nano\second}~\cite{jedec2017ddr4}), assuming a memory controller clock frequency of \SI{3}{\giga\hertz}.}}
2) 4 bits to store the bank id, and 3) 2 bits to store the refresh {type} (Periodic, Preventive, or Invalid). {Our analysis shows that Refresh Table consumes \emph{only} \SI{0.00031}{\milli\meter\squared} chip area per rank and can be accessed in}
\SI{0.07}{\nano\second}.

\noindent
{\textbf{RefPtr Table.} We model a 2048-entry RefPtr Table (128 entries per bank and 16 banks {per} rank). We assume that there can be as many as 1024 rows in a DRAM subarray. Thus, each RefPtr Table entry contains 10 bits to point to a row in a subarray.} {Based on our analysis,} {RefPtr Table{'s size is \SI{0.00683}{\milli\meter\squared} chip area per rank and {it} can be accessed in} \SI{0.12}{\nano\second}}.

\noindent
\textbf{PR-FIFO.}
We model a {4}-entry PR-FIFO per DRAM bank, {assuming the worst case, where the RowHammer defense mechanism generates a preventive refresh for {every} performed row activation}.
{PR-FIFO's chip area cost is \SI{0.00029}{\milli\meter\squared} per rank and a}{ccess {latency is} \SI{0.07}{\nano\second}.}

\noindent
\textbf{Subarray Pairs Table (SPT).} 
{For 128 subarrays per bank, our} analysis shows that this table can be accessed in \SI{0.09}{\nano\second} and consumes {only \SI{0.0018}{\milli\metre\squared}} {chip area} {per DRAM rank.}

\subsection{HiRA-MC's Overall Area Overhead}
\label{sec:hirasched-overall}
{\gls{hirasched} takes only {$0.00923mm^2$} chip area {per DRAM rank}. This area corresponds to \SI{0.0023}{\percent} of the chip area of a \SI{22}{\nano\meter} Intel processor~\cite{inteli7}.}

\subsection{{HiRA-MC's Overall Access Latency}}
\label{sec:hirasched-access-latency}
In the worst-case, \gls{hirasched} traverses the Refresh Table to search for refresh-access parallelization opportunities. {During traversal, within a \gls{trp} time window,} \gls{hirasched} accesses the Refresh Table and {the} SPT 6{8}~times {to iterate over all Refresh Table entries. {{Iterating over} {Refresh Table and SPT {entries}} in a pipelined manner results in an {overall} latency of} {\SI{6.19}{\nano\second}}. {If \gls{hirasched} finds a periodic refresh request, it accesses} RefPtr Table once to {get the address of the row that needs to be refreshed}
{(see \secref{sec:finding-concurrent-refresh})}}{, which takes \SI{0.12}{\nano\second}}. {If \gls{hirasched} finds a preventive refresh, it accesses the head of the PR-FIFO, which takes \SI{0.07}{\nano\second}.}
{Therefore}, the overall access latency of \gls{hirasched} is \SI{6.31}{\nano\second}, {which is significantly smaller than the nominal \gls{trp} of \SI{14.5}{\nano\second}.}
{W}e conclude that {\gls{hirasched} completes all search operations {with} a significantly smaller latency than {the latency of} a precharge operation,} 
and thus it does \emph{not} {cause} additional latency for memory {access} requests.

\setstretch{0.95}
\section{Evaluation Methodology}
\label{sec:usecases}

We evaluate \gls{hirasched} {via} two case studies {focusing on high-density DRAM chips}: 1)~refreshing very high capacity DRAM chips and 2)~protecting DRAM chips with {high} RowHammer vulnerability. We demonstrate for each {study} that \gls{hirasched} {significantly} improves system performance by {leveraging \gls{hira}{'s} {ability to} {concurrently refresh a row {while} refreshing or activating another row.}}

\noindent
\textbf{{Simulation Environment}.}
To evaluate the performance impact of \gls{hirasched} under each use-case, we conduct {cycle-level simulations}, using Ramulator~\cite{ramulatorgithub, Kim2016Ramulator}.
Our baseline leverages the regular {rank-level} \gls{ref} commands, periodically issued at every \gls{trefi} {with a latency of \gls{trfc} in respect to} DDR4 specifications~\cite{jedec2017ddr4}.
Table~\ref{table:system_configuration} {shows the simulated system configuration.} 
In our evaluations, we assume a realistic system with \ncores{} cores, connected to a memory rank with four bank groups each of which contains four banks (16~banks in total). 
{We execute \wlcnt{} 8-core multiprogrammed workloads, randomly chosen from SPEC CPU2006~\cite{spec2006} benchmarks. We simulate these
workloads until each core executes 200M instructions with a warmup period of 100M instructions, similar to prior work~\cite{kim2020revisiting, yaglikci2021blockhammer}.}
The {memory controller} employs {the} FR-FCFS~\cite{rixner00, zuravleff1997controller} scheduling algorithm with {the} open-{row} policy. 
{We assume that a refresh to a DRAM row can be served concurrently with a refresh or an access to \covref{} of the rows within the same DRAM bank,}
based on our experimental results {(\secref{sec:experiments_coverage})}. We {measure system performance in terms of} weighted speedup~\cite{eyerman2008systemlevel, snavely2000symbiotic}.

 \newcolumntype{C}[1]{>{\let\newline\\\arraybackslash\hspace{2pt}}m{#1}}
 \begin{table}[ht]
 \vspace{2mm}
 \caption{{Simulated} system configuration}
 \footnotesize
\centering
\resizebox{\columnwidth}{!}{
 \begin{tabular}{l|l}
 \hline
 \multirow{1}{*}{\textbf{Processor}} & 3.2GHz, 8~core, 4-wide issue, {128-entry} {instr.} window\\ \hline
 \textbf{Last-Level Cache} & {64-byte} cache line, 8-way {set-associative, 8MB} \\ \hline
 \multirow{3}{*}{\textbf{Memory Controller}} & 64-entry each read and write request queues\\&Scheduling policy: FR-FCFS~\cite{rixner00, zuravleff1997controller}\\& Address mapping: MOP~\cite{kaseridis2011minimalistic} \\ \hline
 \multirow{2}{*}{\textbf{Main Memory}} & DDR4-2400~\cite{jedec2017ddr4}, 1 channel{$^*$}, 1 rank{$^*$}, 4 bank groups\\&4 banks/bank group {(16 banks per rank)}, {64K} rows/bank\\ \hline
 \multirow{2}{*}{\textbf{Timing Parameters}} & $t_1=t_2=3ns$, $t_{RC}=46.25ns$, $t_{FAW}=16ns$\\& $ {t_{RefSlack}} \in \{0, 2t_{RC}, 4t_{RC}, 8t_{RC}\}$ \\ \hline\hline
 \end{tabular}
 }
 \begin{flushleft}
 \vspace{-0.5em}
 {$^*$\secref{sec:doduoref} and~\secref{sec:doduopara} assume a 1-channel 1-rank system. \secref{sec:sensitivity} presents a sensitivity analysis for 1, 2, 4, and 8 channels / ranks.}
 \end{flushleft}
 \label{table:system_configuration}
 \end{table}
 
 \noindent{\textbf{Adapting Baseline Refresh for High-Capacity DRAM Chips.}}
Across different generations of DRAM protocols~\cite{jedec2020lpddr5, jedec2020ddr5, jedec2017ddr4, jedec2015lpddr4, jedec2015hbm} the minimum and maximum \glsfirst{trefi} does not significantly change, while the standards allow the manufacturer to define the necessary \glsfirst{trfc} value based on the time required to complete a refresh operation. 
{As DRAM capacity increases, more DRAM rows need to be refreshed when a \gls{ref} command is issued, {which increases} \gls{trfc}~\cite{nguyen2018nonblocking}.}
{To estimate \gls{trfc} for a given density, we use the state-of-the-art regression model~\cite{nguyen2018nonblocking} for projecting \gls{trfc} with increased {chip capacity ($C_{chip}$), {as shown in Expression~\ref{exp:trfc}}:}}
\vspace{-1em}

\begin{equation}
\trfc{} = 110 \times C_{chip}^{0.6}
\label{exp:trfc}
\end{equation}

% \pagebreak
\newif\iflmh
\lmhfalse

\section{{Periodic Refresh Results}}
\label{sec:doduoref}

\begin{figure}[!ht]
    \centering
    \includegraphics[width=\linewidth]{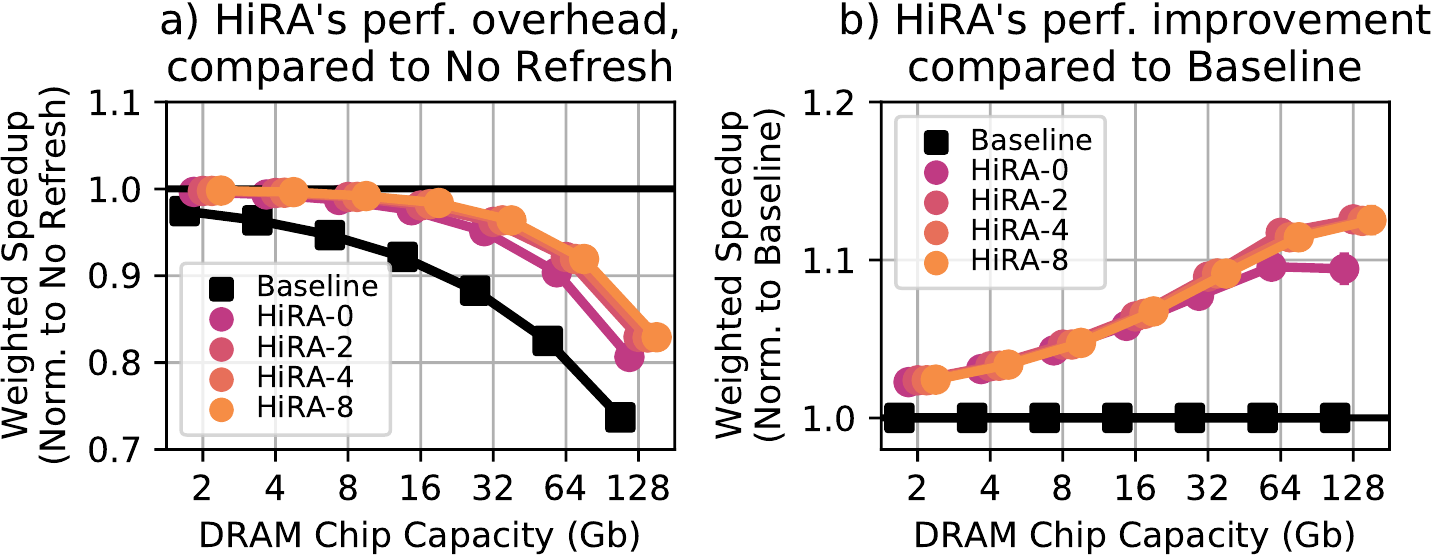}
    \caption{{HiRA's impact on system performance for 8-core multiprogrammed workloads with increasing {DRAM} chip capacity, compared to a)~an ideal system called \emph{No Refresh} that performs \emph{no} periodic refreshes and~b)~{the} Baseline system that uses conventional REF commands to perform periodic refreshes}}
    \label{fig:multi_core_refresh}
\end{figure}

{We evaluate \gls{hira}'s performance when \gls{hira} is used for performing periodic refreshes in high{-}capacity DRAM chips instead of {using conventional} \gls{ref} commands {used in current systems}.}
We sweep DRAM chip {capacity} and {quantify} the performance overhead of periodic refresh operations {{on} {a} baseline {system that performs rank-level \gls{ref} operations} and four configurations of \gls{hira} {with different deadlines}}. 
\figref{fig:multi_core_refresh} demonstrates system performance (y-axis) for different DRAM chip capacities from \SI{2}{Gb} to \SI{128}{Gb} (x-axis).

We denote \gls{hira}'s different configurations as \gls{hira}-N, where N {specifies} {the} {\gls{trrslack} configuration} 
in terms of the number of row activations that can be performed {within a \gls{trrslack}, i.e., \gls{trrslack} of \gls{hira}-N is {$N\times{}\trc{}$}.}

For example, \gls{hira}-2 {schedules} each refresh request with a {\gls{trrslack}} of {$2\times{}\trc{}$}{, whereas \gls{hira}-0 schedules refresh requests with a {\gls{trrslack}} of 0 (i.e., the refresh operation must be performed {immediately after it is generated by \gls{hirasched}})}.
{\figref{fig:multi_core_refresh}a shows system performance in terms of weighted speedup, normalized to an ideal system that we call \emph{No Refresh}, where the system does \emph{not} need to perform any periodic refreshes.}

We make {two} observations {from \figref{fig:multi_core_refresh}a.}
First, {using \gls{ref} commands to perform} periodic refresh operations {({as done in} baseline)} significantly degrades system performance as DRAM chip {capacity} increases. {For example,} periodic refresh operations cause \param{\SI{26.3}{\percent}} system performance degradation for refreshing \param{128Gb} DRAM chips on average across all {evaluated} workloads.
Second, \gls{hira} (\gls{hira}-2) significantly reduces the performance degradation caused by periodic refresh operations by \periodicreduction{} (from {\SI{26.3}{\percent}} down to \param{\SI{17.0}{\percent}}), on average across all {evaluated} workloads {for a DRAM chip capacity of 128Gb}. 

{\figref{fig:multi_core_refresh}b shows system performance in terms of weighted speedup, normalized to the \emph{baseline}. We make three observations from \figref{fig:multi_core_refresh}b.}
{First, \gls{hira} significantly improves system performance. For example, \gls{hira}-2 provides \SI{12.6}{\percent} performance improvement over the baseline on average across all evaluated workloads for a DRAM chip capacity of 128Gb.}
{{Second}, \gls{hira}'s performance benefits increase with \gls{trrslack} {up to a certain {value} of \gls{trrslack}}. For example, for a DRAM chip capacity of 128Gb, \gls{hira}-0 and \gls{hira}-2 provide \SI{9.4}{\percent} and \SI{12.6}{\percent} performance improvement over {the} baseline, respectively.} This is because as \gls{trrslack} increases, \gls{hirasched} can find more opportunities to perform each queued refresh operation concurrently with refreshing or accessing another DRAM row. We observe that increasing \gls{trrslack} from {$2\times{}\trc{}$} to $8\times{}\trc{}$ does \emph{not} significantly improve system performance (i.e., curves for \gls{hira}-2, \gls{hira}-4, \gls{hira}-8 overlap with each other) on average across all {evaluated} workloads. This is because a \gls{trrslack} of {$2\times{}\trc{}$} is large enough to {perform periodic refreshes concurrently with memory accesses or other refreshes}.
{{Third},} \gls{hira}'s performance improvement increases with DRAM chip capacity. For example, \gls{hira}-2's performance improvement increases from \SI{2.4}{\percent} for 2Gb chips to \SI{12.6}{\percent} for 128Gb chips on average across all {evaluated} workloads.

{Based on these observations, we conclude that \gls{hira} significantly improves system performance by reducing the performance overhead of periodic refresh operations{, and HiRA's benefits increase with DRAM chip capacity.}}

\section{RowHammer {Preventive Refresh Results}}
\label{sec:doduopara}

Modern DRAM chips, including the ones that are marketed as RowHammer{-safe~\rhsafe{}}, are shown to be even more vulnerable to RowHammer {(at the circuit level)} than their predecessors~\rhworse{}. Therefore, {it is critical for a RowHammer defense mechanism to efficiently scale} with worsening RowHammer vulnerability. Among {many} RowHammer defense mechanisms {(e.g.,~\rhdef{})}, we {find} \emph{Probabilistic Row Activation (PARA)}~\cite{kim2014flipping} as the most lightweight and {hardware-}scalable RowHammer defense {due to} two reasons.
First, PARA{'s hardware cost does \emph{not} increase when {it is} scaled {to work on chips that have} higher RowHammer vulnerability{. This is} because PARA} {is a \emph{stateless} mechanism {that} refreshes a} potential victim row with a low probability, {defined as} \gls{pth}, when a DRAM row is activated{,} {with \emph{no} need for maintaining {any} metadata}.
Second, PARA{,
as a memory controller-based mechanism which is implemented solely in the processor chip,} easily adapts to the RowHammer {vulnerability} of a {given} DRAM chip by programming \gls{pth} after {the processor is} {deployed}. {Unlike PARA,} other defenses {(e.g.,~\rhdef{})} are {usually} configured for a particular RowHammer vulnerability {level} at {the} processor chip's design time and {they} cannot be easily reconfigured for {a {new} DRAM chip's} {RowHammer} vulnerability. {This is because these mechanisms require implementing as many {hardware} counters as needed to accurately identify a RowHammer attack for a given RowHammer threshold, and thus they {likely} need more counters to reliably work for smaller RowHammer threshold{s}{{; u}nfortunately, the number of {hardware} counters} cannot be {easily} {increased} after deployment}. 

{When scaled {to work on a DRAM chip {that has a} higher} RowHammer vulnerability, PARA refreshes a victim row with a {higher} probability, {thereby} inducing a larger {system} performance overhead~{\cite{kim2020revisiting,yaglikci2021blockhammer}}. \gls{hirasched} {reduces PARA's} {performance overhead}, by leveraging the parallelism \gls{hira} provides. \secref{sec:security_analysis} explains how we configure \gls{pth} when PARA is used with \gls{hira}. Then, \secref{sec:uctwo_eval} evaluates {\gls{hira}'s performance when it is used for performing PARA's preventive refreshes.}}

\subsection{Security Analysis} 
\label{sec:security_analysis}

{\subsubsection{Threat Model}
We assume a comprehensive {RowHammer} threat model,\cqlabel{CQ4} similar to {that assumed by} state-of-the-art works~\cite{kim2014flipping, park2020graphene, yaglikci2021blockhammer}, in which the attacker can 1)~fully utilize memory bandwidth, 2)~precisely time each memory request, and 3)~comprehensively and accurately know details of the memory controller and the DRAM chip. We do not consider any hardware or software component to be trusted or safe except {we assume that the DRAM commands issued by the memory controller are performed within the DRAM chip as intended.}}

\subsubsection{Revisiting PARA's configuration methodology}
\label{sec:revisitingpara}
Kim et al.~\cite{kim2014flipping} {configure} PARA's probability threshold (\gls{pth}), assuming {that} the attacker hammers an aggressor row {\emph{only enough times, but no more}}.
{With} more than an order of magnitude decrease in {the} {RowHammer threshold} in the last decade\cite{kim2020revisiting, frigo2020trrespass, hassan2021utrr, orosa2021deeper, yaglikci2022understanding}, an attacker can {complete a RowHammer attack 144 times {within a refresh window of \SI{64}{\milli\second}},\footnote{{\Gls{hcfirst} for modern DRAM chips is as low as 9600~\cite{kim2020revisiting}. Assuming a \gls{trc} of \SI{46.25}{\nano\second}, performing 9600 row activations can be completed \emph{only} in \SI{444}{\micro\second}, which is 1/144.14 of a nominal refresh window of \SI{64}{\milli\second}. {As such, an attacker can perform 9600 activations for 144 times within a \SI{64}{\milli\second} refresh window.}}}}
{requiring {a} revisit {of PARA's configuration methodology}.\footnote{A concurrent work also revisits PARA's configuration methodology~\cite{saroiu2022configure}.}}

\begin{figure}[!ht]
    \centering
    \includegraphics[width=0.8\linewidth]{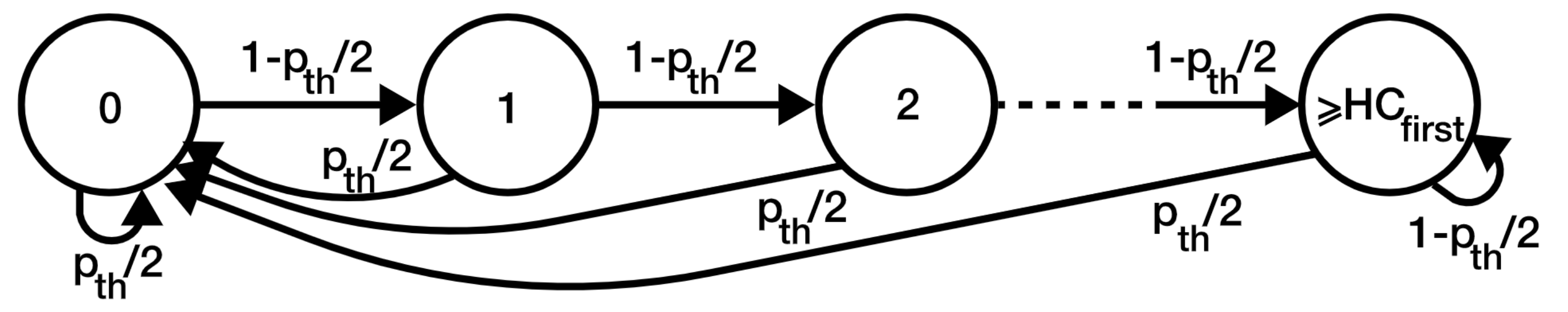}
    \caption{Probabilistic state machine of {hammer} count in a PARA-protected system}    \label{fig:para_fsm}
\end{figure}

{The rest of this section explains how we {calculate}} \gls{pth} for a given RowHammer threshold.
\figref{fig:para_fsm} shows the probabilistic state machine that we use to 
calculate the {hammer count, which we define as the number of aggressor row activations} that may affect a victim row.Initially the {hammer count} is zero (state~0). When an {aggressor row} {is} activat{ed}, PARA {triggers a \emph{preventive} refresh with a probability of $\pth{}$. Since there are two rows that are adjacent to the activated row, PARA} refreshes {the} victim row with a probability of $\pth{}/2$, {in which case the {hammer} count is reset.} {Therefore,} the {hammer} count {{is incremented} with a probability of $1-\pth{}/2$ upon an {aggressor row} activation}. An attack is considered to be \emph{successful} if {its hammer count} reaches the RowHammer threshold {($N_{RH}$)} within a \glsfirst{trefw}. 

{Because the time delay between two row activations targeting the same {bank} \emph{cannot} be smaller than \gls{trc}, an} attacker can perform {a maximum of} $\trefw{}/\trc{}$ state transitions within a \gls{trefw}. 
To account for all possible access patterns, we model a \emph{successful RowHammer {access pattern}} as a set of failed attempts{, where} the victim row is refreshed before the {hammer} count reaches {the} RowHammer threshold, followed by a successful attempt, {where} the victim row is {\emph{not}} refreshed until the {hammer} count reaches {the} RowHammer threshold. {To {calculate} \gls{pth}{,} we follow a {five-}step approach. First, we calculate the probability of a failed attempt ($p_f$) and a successful attempt. Second, we calculate {the probability of observing a number ($N_f$) of consecutive failed attempts.}Third, we calculate
{\emph{the {overall} RowHammer success probability ($p_{RH}$)} as the overall probability of {\emph{all} possible successful RowHammer access patterns.}}
Fourth, we extend the probability calculation to account for \gls{trrslack}. Fifth, we calculate \gls{pth} for a given failure probability target.}

\pagebreak
\head{{Step {1}: Failed and successful attempts}}
\expref{exp:failedattempt} {shows} $p_{f}(HC)$: the probability of a {\emph{failed} attempt with a given hammer count ($HC$). The attempt contains 1)~$HC$ consecutive aggressor row activations that do \emph{not} trigger a preventive} refresh, $(1-\pth{}/2)^{HC}$, {where $HC$ is smaller than the RowHammer threshold ($N_{RH}$)}, {and 2)~}an {aggressor row} activation {that triggers} a {preventive} refresh $(\pth{}/2)$.

\equ{p_{f}(HC) = (1-\pth{}/2)^{HC}\times\pth{}/2,~where~1\leq HC<N_{RH}}{exp:failedattempt}

{Similarly, we calculate the probability of a \emph{successful} attempt which has $N_{RH}$ consecutive aggressor row activations that do \emph{not} trigger a preventive refresh {as} $(1-p_{th}/2)^{N_{RH}}$.}

\head{{Step {2}: {The probability of} ${\bm{N_f}}$ {consecutive} failed attempts}}
{{Since a failed attempt may have a hammer count value in the range $[1, N_{RH})$, we account} for all possible hammer count {values} that {a failed attempt might have}.}
{\expref{exp:nconsecutivefailedattempts} shows the probability of {a given number of (}$N_f${)} consecutive failed attempts.}
\equ{\prod_{i=1}^{N_{f}} p_f(HC_i)=\prod_{i=1}^{N_{f}}((1-\pth{}/2)^{HC_i}\times\pth{}/2)~, where~1\leq HC_{i} < \hcfirst{}}{exp:nconsecutivefailedattempts}

\head{{Step 3: {Overall} RowHammer success probability}}
{To find the overall RowHammer success probability, we 1)~calculate the probability of the \emph{successful RowHammer access pattern {($p_{success}(N_{f})$),} which consists of $N_f$ consecutive failed attempts and one successful attempt} for each possible value that $N_f$ can take and 2)~sum the probability of all possible successful RowHammer access patterns: $\sum{p_{success}(N_f)}$.}
To do so, {we} multiply \expref{exp:nconsecutivefailedattempts} with the probability of a successful attempt: $(1-p_{th}/2)^{N_{RH}}$.
{\expref{exp:successfulattack}} shows {how we calculate $p_{success}(N_{f})$.} We derive \expref{exp:successfulattack} by evaluating the product on both terms in \expref{exp:nconsecutivefailedattempts}: $p_{th}/2$ and $(1-p_{th}/2)^{HC_i}$.

\equ{p_{success}(N_{f}) = (1-{\pth{}}/{2})^{\sum_{i=1}^{N_{f}} {HC_{i}}} \times ({\pth{}}/{2})^{N_{f}} \times (1-{\pth{}}/{2})^{\hcfirst{}}}{exp:successfulattack}

To account for the worst-case, we maximize 
$p_{success}(N_{f})$ {by choosing} the worst possible value for each $HC_{i}$ {value}. \textbf{Intuitively}, {the number of activations in a failed attempt should be minimized. Since} a failed attempt {has to perform at least one activation,
we conclude that all failed attempts} fail after only one row activation in the worst case. 
\textbf{Mathematically,} {our goal is} to maximize $p_{success}(N_{f})$. {{Since} $\pth{}$ is a value between zero and one,} we minimize the term $\sum_{i=1}^{N_{f}} {HC_{i}}$ to maximize $p_{success}(N_{f})$. Thus, we derive \expref{exp:prhn} {by choosing $HC_i=1$ to achieve} {the maximum (worst-case) $p_{success}(N_{f})$.}

\equ{
p_{success}(N_{f}) = (1-{\pth{}}/{2})^{N_{f}+\hcfirst{}} \times ({\pth{}}/{2})^{N_{f}}}{exp:prhn}

{{\expref{exp:prh}} shows {{the overall} RowHammer success probability} ($p_{RH}$), as the sum of all possible $p_{success}(N_f)$ values.}
{$N_{f}$ can be as small as 0 if {the} RowHammer attack does \emph{not} include any failed attempt and as large as} the maximum number of failed attempts that can fit in a refresh window {(\gls{trefw}) {together with a successful attempt}}.
Since $HC_i=1$ in the worst case, each failed attempt {costs only two row} activations ($2\times\trc{}$){:} {one aggressor row activation and one activation for preventively refreshing the victim row}. Thus, the execution time of $N_{f}$ {failed} attempts, {followed by one successful attempt is} $(2N_{f}+\hcfirst{})\times\trc{}$. {Therefore,} the maximum {value $N_{f}$ can take within a refresh window (\gls{trefw})} is ${N_{f}}_{max} = ((\trefw{}/\trc{})-\hcfirst{})/2$.

\equ{p_{RH} = \sum_{N_{f}=0}^{{N_{f}}_{max}} p_{success}(N_{f}),~~~{N_{f}}_{max} = (\trefw{}/\trc{}-\hcfirst{})/2}{exp:prh}

Using {\expsref{exp:prhn}} and~\ref{exp:prh}, we compute the {{overall} RowHammer success probability} for a given {PARA} probability threshold ($p_{th}$).

\head{{Step 4: Accounting for \gls{trrslack}}}
The original PARA proposal~\cite{kim2014flipping} performs a {preventive} refresh immediately after the activated row is closed. However, \gls{hirasched} allows a {preventive} refresh to be queued {for a time window as long as \gls{trrslack}}. {Since the aggressor rows can be activated while a preventive refresh request is queued, we update {~\expsref{exp:prhn}} and~\ref{exp:prh}, assuming the worst case, where an aggressor row is activated as many times as possible {within \gls{trrslack} ({i.e.,} the maximum amount of time} the preventive refresh is queued).}
To do so, we {update} \expref{exp:prh}: {we reduce} {the RowHammer threshold ($N_{RH}$)} by {\gls{hcdeadline}}. Thus, we calculate ${N_{f}}_{max}$ and $\prh{}$ as shown in \expsref{exp:nmax} and~\ref{exp:prhdl}, respectively. 

\begin{equation}\footnotesize
{N_{f}}_{max} = ((\trefw{}/\trc{})-\hcfirst{}-{\hcdeadline{}})/2
\label{exp:nmax}
\end{equation}

\equ{p_{RH} = \sum_{N_{f}=0}^{{N_{f}}_{max}} (1-{\pth{}}/{2})^{{N_{f}}+\hcfirst{}-{\hcdeadline{}}} \times ({\pth{}}/{2})^{N_{f}}}{exp:prhdl}

\head{{Step 5: Finding $\bf{p_{th}}$}}
{We iteratively evaluate \expsref{exp:nmax} and \ref{exp:prhdl} to find} \gls{pth} for a {target} {{overall} RowHammer success probability of {$10^{-15}$}, as a typical consumer memory reliability target {(see, e.g.,~\cite{cai2012error, cai2017flashtbd, jedec2011failure, luo2016enabling, patel2017reaper, yaglikci2021blockhammer, kim2020revisiting, micheloni2015apparatus})}}.

\subsubsection{{Results}}

{We refer to {the} original PARA work~\cite{kim2014flipping} as PARA-Legacy. PARA-Legacy calculates {{the overall} RowHammer success probability} as ${p_{RH_{Legacy}}=(1-p_{th}/2)^{N_{RH}}}$ with an optimistic assumption that the attacker hammers an aggressor row \emph{only enough times, but no more}. To mathematically compare {{the overall} RowHammer success probability} that we calculate ($p_{RH}$) with $p_{RH_{Legacy}}$, we reorganize \expref{exp:prhdl}, which already includes ${p_{RH_{Legacy}}=(1-p_{th}/2)^{N_{RH}}}$, and derive \expref{exp:prhdiff}.}

\equ{
p_{RH} =k \times p_{RH_{Legacy}},~where~k = (1-\frac{\pth{}}{2})^{-{\hcdeadline}}\sum_{N_{f}=0}^{{N_{f}}_{max}}(\frac{\pth{}}{2}(1-\frac{\pth{}}{2}))^{{N_{f}}}
}{exp:prhdiff}

{
\expref{exp:prhdiff} shows that $p_{RH}$ {is a multiple of} $p_{RH_{Legacy}}$ by a factor of $k$, {where $k$ depends on a given system's properties (e.g., $N_{f_{max}}$)} and PARA's configuration (e.g., \gls{pth}).
{To understand the difference between $p_{RH}$ and $p_{RH_{Legacy}}$, we evaluate \expref{exp:prhdiff} for different RowHammer threshold values.\footnote{{We calculate for $N_{RefSlack}=0$, $t_{REFW}=64ms$, and $t_{RC}=46.25ns$.}}}
For {old DRAM chips (manufactured in} 2010-2013)~\cite{kim2014flipping}, $k$ is 1.0005 {(for $N_{RH}=50K$ and $p_{th}=0.001$~\cite{kim2014flipping})}, causing \emph{only} \SI{0.05}{\percent} variation in PARA's reliability target. {However,for future DRAM chips with $N_{RH}$ values of 1024 and 64 {($p_{th}$ values of 0.4730 and 0.8341)}, $k$ becomes 1.0331 and 1.3212, respectively. Therefore, the difference between {the two probabilities, $p_{RH}$ and $p_{RH_{Legacy}}$,} significantly increases as RowHammer worsens.}}

\begin{figure}[!ht]
    \centering
    \includegraphics[width=\linewidth]{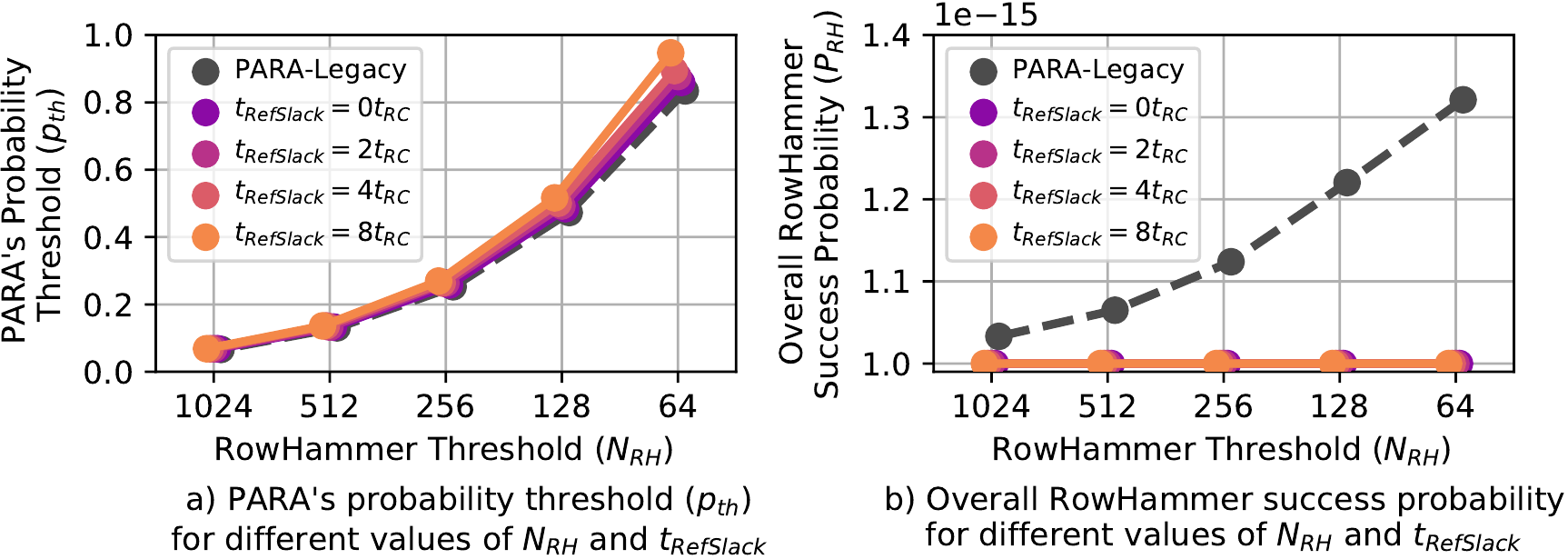}
    \caption{{PARA configurations for different RowHammer thresholds ($N_{RH}$) and \gls{trrslack} values: a)~PARA's probability threshold ($p_{th}$}) and b)~{{overall} RowHammer success probability} ($p_{RH}$)}
    \label{fig:para_thresholds}
\end{figure}

\figref{fig:para_thresholds}{a} show{s} how {decreasing RowHammer threshold, i.e., worsening RowHammer vulnerability {(x-axis)}, change{s}} {PARA's {probability} threshold (\gls{pth}) {(y-axis)}}. {The dashed curve shows} PARA-Legacy's probability threshold (calculated using ${p_{RH_{Legacy}}=(1-p_{th}/2)^{N_{RH}}}$), {whereas the other curves show $p_{th}$ for different \gls{trrslack}} {values which we} {calculate} using {\expref{exp:prhdl}}.

{We make two observations from \figref{fig:para_thresholds}a.} First, to maintain {a $10^{-15}$} RowHammer {success} {probability}, $p_{th}$ significantly increases for smaller RowHammer thresholds.
{For example,} $p_{th}$ {increases} from 0.068 to 0.860 {(\gls{trrslack}=0)} when the RowHammer threshold reduces from 1024 to 64. This is because as the RowHammer threshold reduces, fewer activations {are} enough for an attack to induce bit flips. {Thus,} PARA needs to perform preventive refreshes more aggressively.
{Second,}
$p_{th}$ {increases with} \gls{trrslack}, e.g., when {the} RowHammer threshold is 128, $p_{th}$ should be 0.48, 0.49, 0.50, and 0.52 for {\gls{trrslack} values of 0, 2\gls{trc}, 4\gls{trc}, and 8\gls{trc}, respectively.} 
This is because a larger {\gls{trrslack} {allows reaching a higher hammer count, requiring PARA to perform preventive refreshes more aggressively.}} 

\figref{fig:para_thresholds}{b} show{s} how {decreasing RowHammer threshold ($N_{RH}$) change{s}} {the {{overall} RowHammer success probability} ($p_{RH}$).} {We calculate \emph{all} $p_{RH}$ values by evaluating \expref{exp:prhdl} using the $p_{th}$ values in \figref{fig:para_thresholds}a.} {The dashed curve shows} PARA-Legacy's $p_{RH}$, {whereas the other curves show PARA's $p_{RH}$ for different \gls{trrslack} configurations.}
{We make two observations from \figref{fig:para_thresholds}b.}
{{First}, configuring $p_{th}$ as described in PARA-Legacy~\cite{kim2014flipping} {1)~}results in a larger {{overall} RowHammer success probability} than {the consumer memory reliability target ($10^{-15}$)}, and {2)~}the difference between $p_{RH}$ and the $10^{-15}$ increases as the RowHammer vulnerability increases (i.e., the RowHammer threshold reduces). For example, the $p_{th}$ values {that PARA-Legacy calculates}
{targeting a $10^{-15}$ {overall} RowHammer success probability} {for RowHammer thresholds of 1024 and 64,} result in {{overall} RowHammer success probability {values} of} {$1.03\times{}10^{-15}$} and {$1.32\times{}10^{-15}$}, {respectively.}
This happens because PARA-Legacy assumes that the attacker performs \emph{{only}} as many {aggressor row} activations as the RowHammer threshold ($N_{RH}$) within a refresh window, even though {increasingly} more {aggressor} row activations can {be performed} in a refresh window {as $N_{RH}$ reduces}.}
{Second}, {the} {$p_{th}$ values} {that we calculate using \expref{exp:prhdl}} significantly reduce the {{overall} RowHammer success probability} {compared to PARA-Legacy} {(and maintains a $p_{RH}$ of $10^{-15}$ across all RowHammer thresholds)} {because \expref{exp:prhdl}, {to calculate $p_{RH}$,} takes into account all aggressor row activations that can be performed in a refresh window.}

{{W}e conclude that as RowHammer threshold decreases, PARA-Legacy{'s $p_{th}$ values} result in a significantly larger {{overall} RowHammer success probability} than the {consumer memory reliability target ($10^{-15}$)}, while $p_{th}$ {values} calculated using \expref{exp:prhdl} maintain the {{{overall} RowHammer success probability} at $10^{-15}$}. }

\subsection{Performance of PARA with HiRA}
\label{sec:uctwo_eval}
{We evaluate \gls{hira}'s performance {benefits} when it is used {to} perform PARA's preventive refreshes. {We} us{e} the evaluation methodology described in \secref{sec:usecases}. {We calculate \gls{pth} values \expref{exp:prhdl}.}
{{W}e evaluate PARA's {impact on} performance when it is used \emph{with} and \emph{without} \gls{hira} for different RowHammer thresholds}.}

\begin{figure}[!ht]
    \centering
    \includegraphics[width=\linewidth]{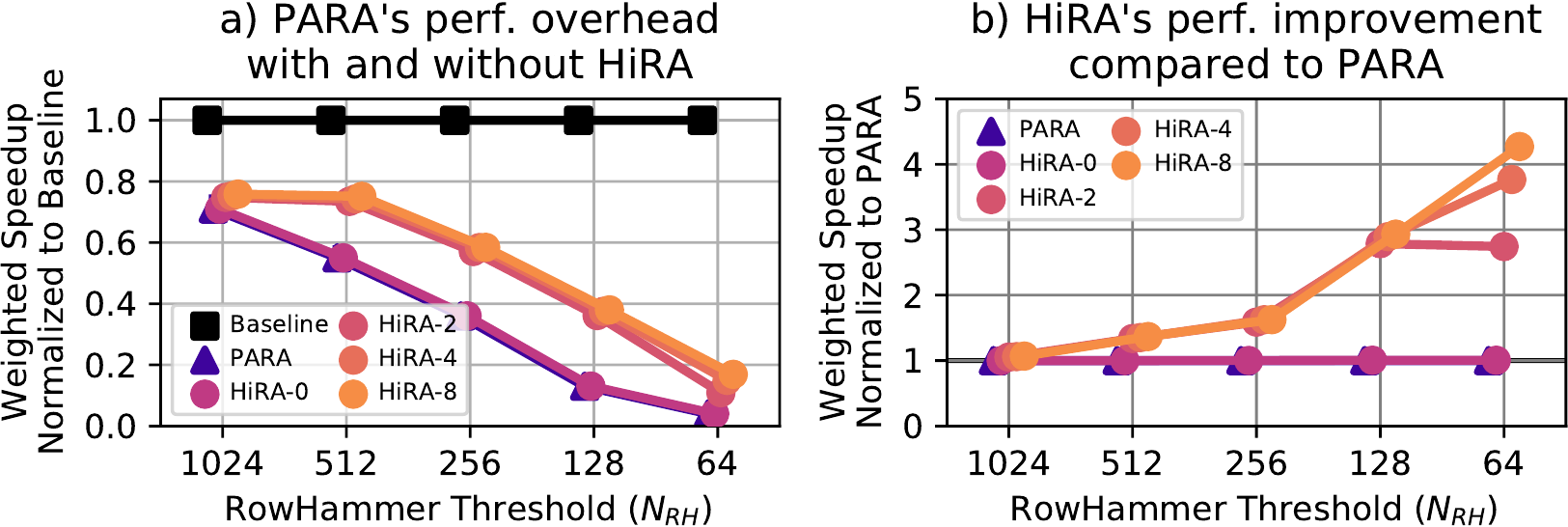}
    \caption{\gls{hira}'s {impact on} system performance {for 8-core multiprogrammed workloads with increasing RowHammer vulnerability (i.e., {decreasing} $N_{RH}$) compared to a) Baseline system with \emph{no} RowHammer defense and b) a system that implements PARA.}}
    \label{fig:para_perf_multicore}
\end{figure}

\figref{fig:para_perf_multicore}{a} shows {the performance of 1)~a system that implements PARA without \gls{hira}, labeled as PARA, and 2)~a system that implements PARA with {four different \gls{trrslack} configurations of} \gls{hira}, labeled {using the HiRA-N notation (\gls{trrslack} of HiRA-N is $N\times{}\trc{}$)} as HiRA-{0, HiRA-2, HiRA-4, and HiRA-8}, {normalized to {the} baseline that does \emph{not} perform any preventive refresh operations (i.e., does \emph{not} implement PARA).}}

{From \figref{fig:para_perf_multicore}a, we observe that}
PARA induces {\SI{29.0}{\percent}} slowdown on system performance on average across all {evaluated} workloads when it is configured for a RowHammer threshold of 1024. \gls{hira}-2 reduces PARA's performance overhead down to {\SI{25.2}{\percent}}, which results in a performance improvement of {\SI{5.4}{\percent}} compared to PARA.
Similarly,  when configured for a RowHammer threshold of 64, \gls{hira}-4 {increases system performance by $3.73\times$ compared to PARA {as it} reduc{es} PARA's performance overhead by \reactivereduction{} (from} \SI{96.0}{\percent} down to \SI{85.1}{\percent}{)}. 
{This happens because HiRA reduces the latency of preventive refreshes by concurrently performing them with refreshing or accessing other rows in the same bank.}

{\figref{fig:para_perf_multicore}b shows the performance of the system that implements PARA with \gls{hira} (labeled using the HiRA-N notation), normalized to the performance of the system that implements PARA without \gls{hira} (labeled as PARA).}
{We make two observations from \figref{fig:para_perf_multicore}b.}
{First, \gls{hira}'s performance improvement increases with higher RowHammer vulnerability, i.e., smaller RowHammer threshold.}
{For example, when compared to PARA (\figref{fig:para_perf_multicore}b), \gls{hira}-2 provides a speedup of $2.75\times$ on average across all {evaluated} workloads when RowHammer threshold is 64, which is significantly larger than \gls{hira}-2's performance improvement of \SI{5.4}{\percent} when RowHammer threshold is 1024.}
{This is because PARA generates preventive refreshes more aggressively as RowHammer threshold reduces (\secref{sec:revisitingpara}), which increases PARA's memory bandwidth utilization and provid{es} HiRA with a larger number of preventive refreshes to parallelize with other accesses and refreshes.}
{Second, configuring \gls{hira} with a larger \gls{trrslack} improves system performance. For example, when {the} RowHammer threshold is \SI{64}{\milli\second}, \gls{hira}-0, \gls{hira}-2, \gls{hira}-4, and \gls{hira}-8 improve system performance by \SI{0.6}{\percent}, $2.75\times$, $3.73\times$, and $4.23\times$, respectively, on average across all {evaluated} workloads, compared to PARA without \gls{hira} (\figref{fig:para_perf_multicore}b). This happens because \gls{hirasched} can find a parallelization opportunity for a queued preventive refresh with a larger probability when there is a larger \gls{trrslack}.}

{Based on {our} observations, we conclude that \gls{hira} significantly reduces PARA's system performance degradation.}

\setstretch{0.9425}
\section{\sho{Sensitivity Studies}}
\label{sec:sensitivity}
\sho{We analyze how HiRA's performance changes {with} 1)~number of channels and 2)~number of ranks per channel. To {evaluate} high-end system configurations, we sweep the number of channels and ranks from one to eight, {inspired by} commodity systems~\cite{intelxeon, amdepyc7003, micronnaming, jedec2017ddr4, micron2014ddr4, micron2014networking}.}

\subsection{HiRA with Periodic Refresh}
\begin{figure}[!ht]
    \centering
    \includegraphics[width=\linewidth]{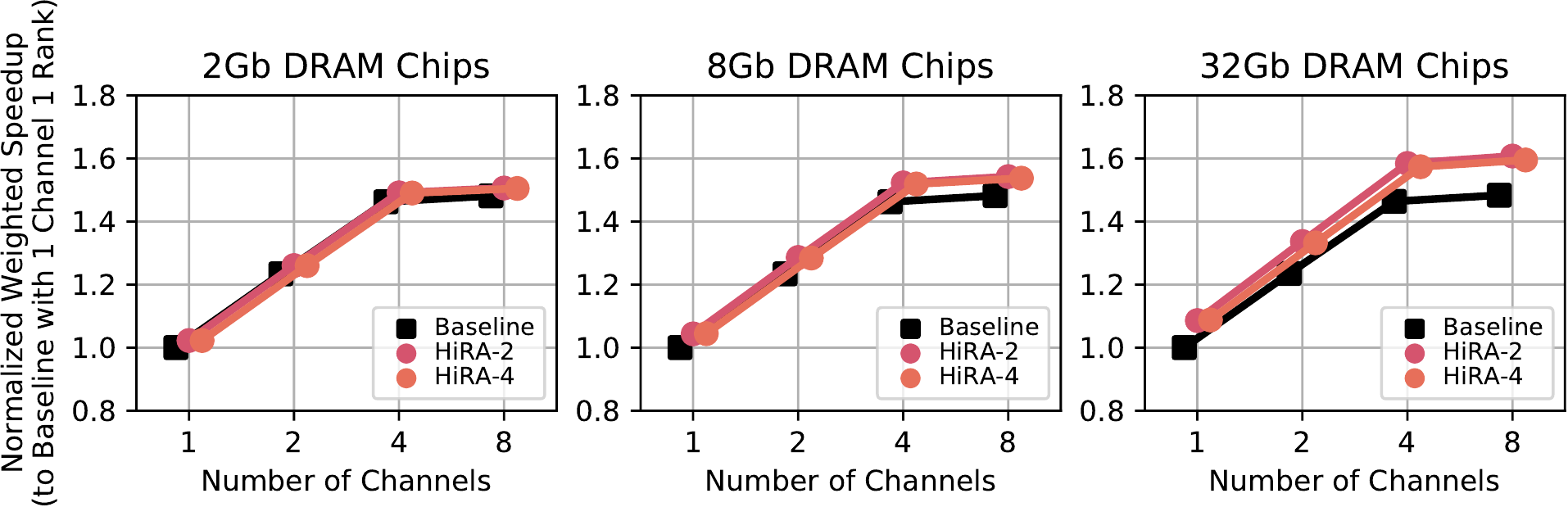}
    \caption{\sho{{E}ffect of \shom{channel count} on system performance for Baseline and HiRA}}
    \label{fig:ref_channel}
\end{figure}

\sho{\figref{fig:ref_channel} shows how increasing \shom{the} number of channels {(x-axis)} affects HiRA's performance for two configurations (HiRA-2 and HiRA-4), compared to the baseline, where rows are periodically refreshed using rank-level $REF$ command. {The y-axis shows system performance in terms of {average} weighted speedup across 125 evaluated workloads,} normalized to the baseline's performance at \shom{the} 1-channel 1-rank configuration.}
{Three subplots show the results for 2Gb (left), 8Gb (middle), and 32Gb (right) {DRAM} chip capacity.}

\sho{We make three observations.
First, both {the} baseline and HiRA {provide} \shom{higher} performance with \shom{more} channels. For example, HiRA and {the} baseline exhibit speedups of $1.60\times$ and $1.48\times$, \shom{respectively,} when the number of channels \shom{increases} from one to eight for a DRAM chip {capacity} of 32Gb. This is because memory-level parallelism {increases with more channels}. {P}erformance overhead\shom{s} of rank-level refresh and HiRA do \emph{not} increase with \shom{more} channels because different channels do not share command, address, or data bus{es}, thereby allowing {different channels} to be \shom{accessed} simultaneously.  
Second, {at smaller channel counts,} the effect of channel count {on performance} is greater. For example, the slope\shom{s} of the line plots are steeper in-between one and four channels than in-between four and eight channels. This happens because {the evaluated} workloads do not exhibit \shom{sufficient} memory-level parallelism to fully leverage the available parallelism \shom{with} {more than} four channels. 
Third, both HiRA-2 and HiRA-4 configurations exhibit {significant} speedup {over} {the} baseline for \emph{all} channel counts. For example, \gls{hira}-2 improves the performance of a system using 32Gb DRAM chips with eight channels by \SI{8.1}{\percent} compared to {the baseline with} 8-channels.} 
{We conclude that \gls{hira} provides significant performance benefits for high-capacity DRAM chips {even with a large number of channels}.}

\begin{figure}[!ht]
    \centering
    \includegraphics[width=\linewidth]{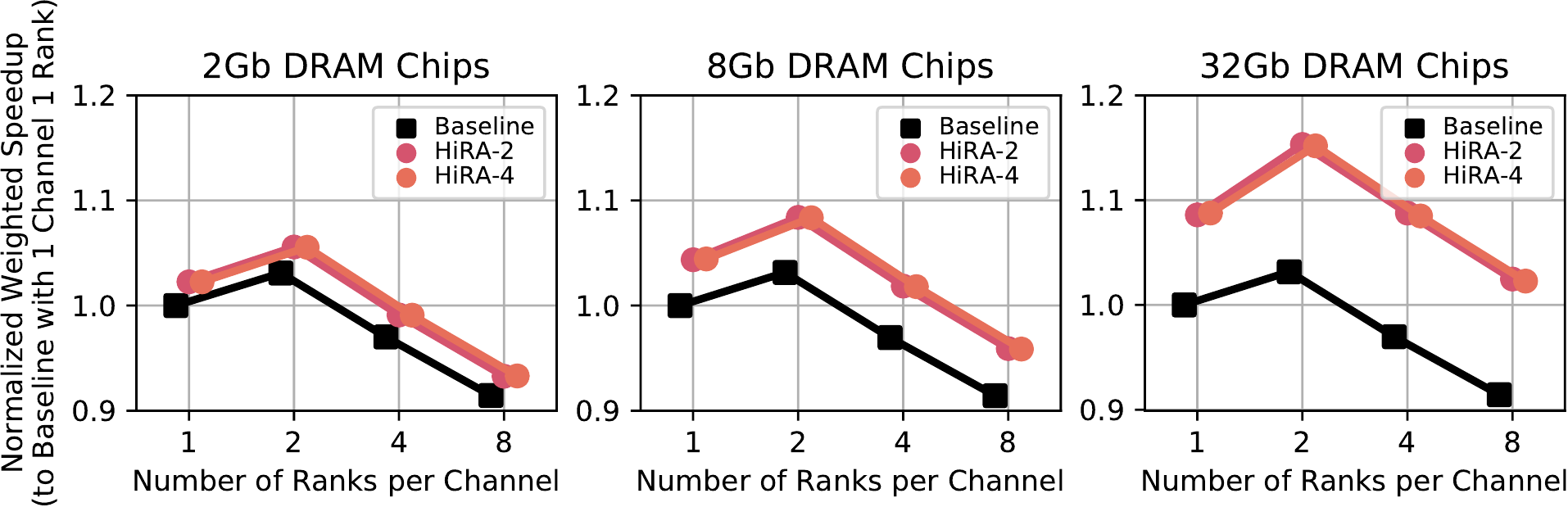}
    \caption{\sho{{E}ffect of {rank count} on system performance for Baseline and HiRA}}
    \label{fig:ref_rank}
\end{figure}

{{\figref{fig:ref_rank}} shows how increasing \shom{the} number of {ranks} (x-axis) affects HiRA's performance benefits. The y-axis shows system performance using the same metric as \figref{fig:ref_channel} uses. {Three subplots show the results for 2Gb (left), 8Gb (middle), and 32Gb (right) {DRAM} chip capacity.}}

\sho{We make three observations.
First, increasing the number of ranks from one to two increases system performance {(e.g., by \SI{3}{\percent} {and} \SI{15.3}{\percent} for {the} baseline {and} \gls{hira}-2, {respectively,} for a chip capacity of 32Gb)}. This is because the {evaluated} workloads leverage the {higher} rank-level parallelismSecond, \shom{unlike {with} channels,} further {increasing} the number of ranks beyond two \emph{slows down} the system for both \shom{the} baseline and HiRA {by \SI{11.7}{\percent} and \SI{11.1}{\percent}, respectively, on average {as} number of ranks increases from 2 to 8}. This happens because multiple ranks share a single command bus and {together} occupy the command bus for refresh operations, {making the command bus a bottleneck}. Third, HiRA {provides} higher performance than \shom{the} baseline for \emph{all} {evaluated} rank configurations.}
{For example, \gls{hira}-2 provides \SI{12.1}{\percent} performance improvement over {the} baseline {even} for an 8-rank system with 32Gb DRAM chips.}
{We conclude that \gls{hira} provides significant performance benefits for high-capacity DRAM chips {compared to the baseline even with a large number of ranks}.}

\subsection{HiRA with Preventive Refresh}

\begin{figure}[!ht]
    \centering
    \includegraphics[width=\linewidth]{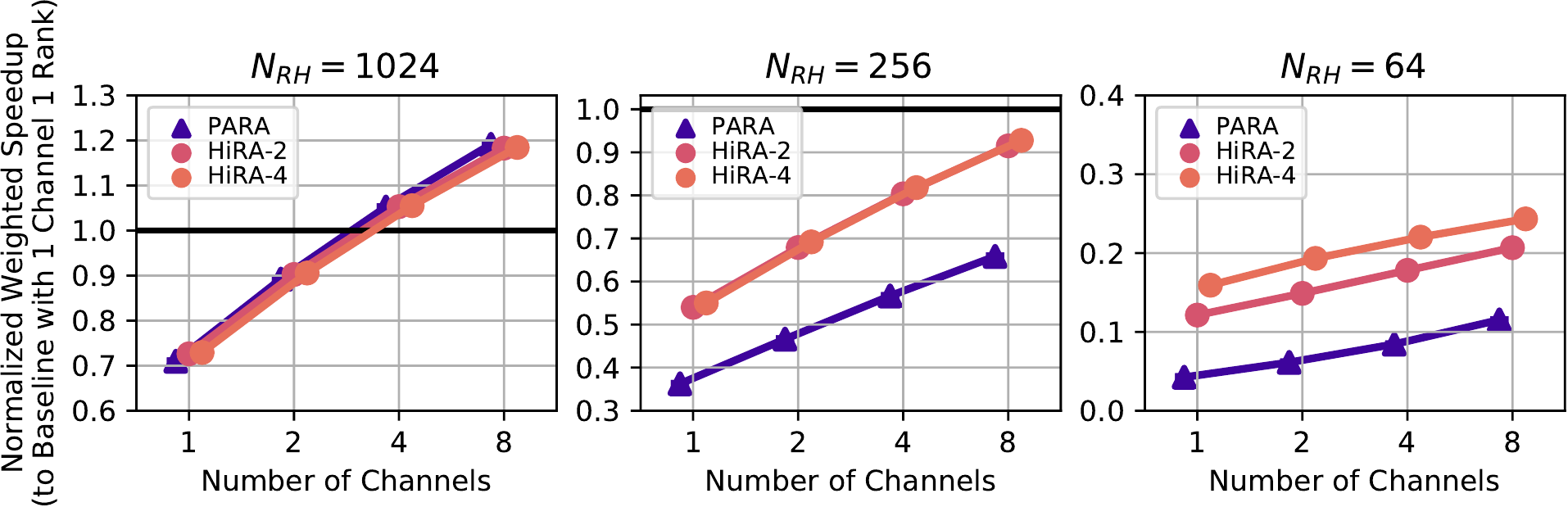}
        \caption{\sho{{E}ffect of \shom{channel count} on system performance {with} PARA and HiRA}}
    \label{fig:para_channel}
\end{figure}

\sho{\figref{fig:para_channel} shows how increasing \shom{the} {channel count} {(x-axis)} affects 
{PARA's impact on system performance when used without {\gls{hira}} (labeled as PARA) and with HiRA (labeled as HiRA-N, where N represents the \gls{trrslack} configuration as in \secref{sec:doduoref} and~\secref{sec:uctwo_eval}).}
{The y-axis reports system performance in terms of average weighted speedup across 125 workloads,} normalized to the baseline 1-channel 1-rank system with \emph{no} RowHammer defense mechanism.
{Three subplots show the results for RowHammer thresholds ($N_{RH}$) of 1024 (left), 256 (middle), and 64 (right).}}

\sho{We make {three} observations.
First, {system} performance increases with \shom{channel count} when \shom{PARA} is used {(with and without HiRA)}. For example, increasing \shom{the} number of channels from one to eight {improves the performance of the system with PARA {and \gls{hira}-2} by \SI{67.6}{\percent}}
{and by \SI{63}{\percent}, respectively,}
when the RowHammer threshold is 1024. This is because memory accesses are distributed across a larger number of banks \shom{given more channels}, thereby reducing the congestion in banks, and thus the \shom{number of} row buffer conflicts. As a result, {the evaluated} workloads perform fewer row activations, and PARA generates fewer {preventive} refreshes.
Second, {at smaller RowHammer thresholds, \gls{hira} significantly improves system performance even with a large number of channels.} 
{For example,}
PARA \shom{{causes}}
{\SI{88.5}{\percent}} performance reduction on an eight-channel system when the RowHammer threshold is 64. HiRA{-2 and HiRA-4 reduce} this performance overhead {to {\SI{79.3}{\percent} and \SI{75.7}{\percent}}, respectively,} by {performing preventive refreshes concurrently with refreshing or activating other rows in the same bank.}
{Third, \gls{hira} improves system performance compared to PARA for all {evaluated} channel counts. This happens because \gls{hira} reduces the performance overhead of PARA's preventive refreshes for each memory channel regardless of the system's channel count.
}
{We conclude that \gls{hira} provides significant performance benefits even with {a} {large number of channels.}}
}

\begin{figure}[!ht]
    \centering
    \includegraphics[width=\linewidth]{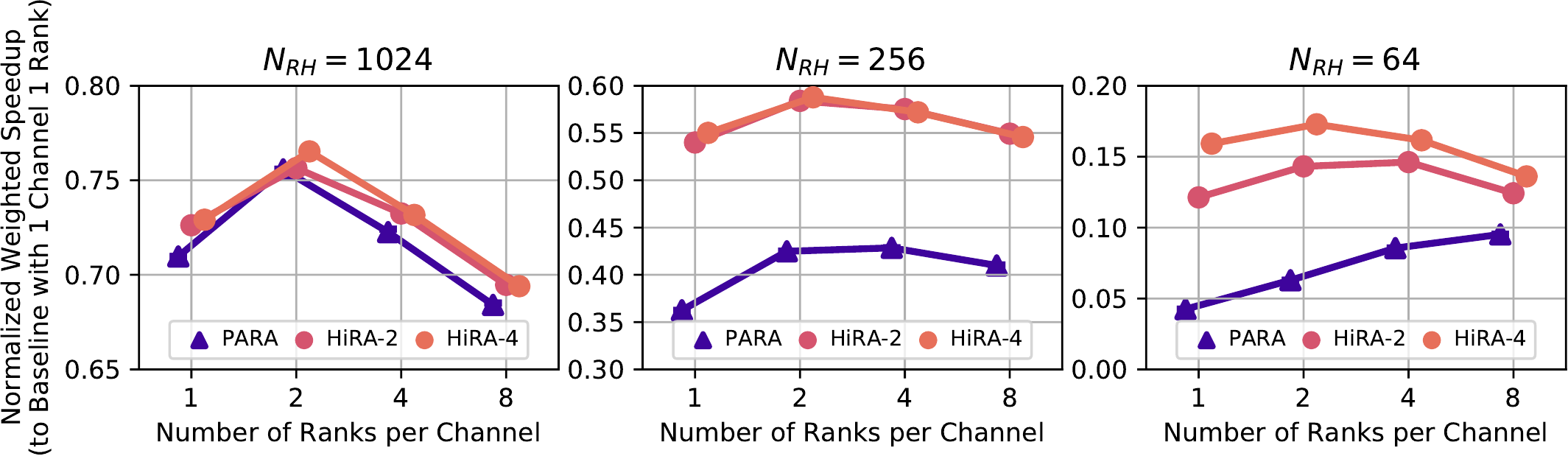}
    \caption{\sho{{E}ffect of \shom{{rank} count} on system performance {with} PARA and HiRA}}
    \label{fig:para_rank}
\end{figure}

{\figref{fig:para_rank} shows how increasing \shom{the} {rank count} {(x-axis)} affects 
{PARA's impact on system performance when used without {\gls{hira}} and with HiRA.}
{The y-axis shows system performance {using the same metric as \figref{fig:para_channel} uses}.
{Three subplots show the results for RowHammer thresholds ($N_{RH}$) of 1024 (left), 256 (middle), and 64 (right).}}}

We make three observations.
First, similar to \figref{fig:ref_rank}, increasing the number of ranks from one to two increases system performance for all three mechanisms {(e.g., by \SI{6.5}{\percent} and \SI{4.9}{\percent} for PARA and HiRA-4 when $N_{RH}=1024$)} across all shown RowHammer thresholds due to {the higher} rank-level parallelism.
Second, further {increasing} the number of ranks beyond two ranks reduces HiRA's benefits over PARA. {{This happens because} increasing the {rank count}
beyond two {increases the command bus bandwidth usage of periodic refresh requests.}{Third, despite the performance reduction at high rank counts, \gls{hira} significantly improves system performance compared to PARA. For example, \gls{hira}-2 (\gls{hira}-4) {improves system performance by} \SI{30.5}{\percent} (\SI{42.9}{\percent}) compared to PARA on an 8-rank system with a RowHammer threshold of 64.}
{Based on these observations, we conclude that \gls{hira} provides significant performance benefits over PARA even when a large number of ranks {share the command bus.}}}

\section{Summary of {Major} Results}
\label{sec:takeaways}
{We summarize} the {major} observations from {four main} evaluations in this paper. 
First, {by using \gls{hira}, it is possible to reliably} refresh a DRAM row {concurrently with}
refreshing or {activating} another DRAM row within the same bank {in} off-the-shelf DRAM chips {(\secref{sec:characterization})}. {\secref{sec:characterization}} experimentally demonstrates {on \chipcnt{} real DRAM chips} that 
{\gls{hira} {can} reliably parallelize a DRAM row's {refresh operation} {with refresh or {activation of} {any of the}}} \covref{} of the rows within the same bank.
Second, \gls{hirasched} reduces the {overall latency of {refreshing} two DRAM rows within the same bank} by \param{\SI{51.4}{\percent}} (\secref{sec:characterization}).
{Third, {\gls{hira} {significantly improves system performance by} reducing} the performance {degradation} {caused by} periodic refreshes across \emph{all} system configurations we evaluate (\secref{sec:doduoref}). 
{Fourth}, \gls{hirasched} {significantly} {improves system} performance {by reducing the performance overhead of PARA's preventive refreshes across \emph{all} system configurations we evaluate (\secref{sec:doduopara})}.
{Our major results show that \gls{hira}}
can {effectively and robustly} improve {system performance} by reducing the time spent for \emph{both} periodic refreshes and {preventive} refreshes without compromising system reliability or security. We hope that our findings inspire DRAM manufacturers and {standards bodies} to {explicitly and} {properly} support \gls{hira} in future DRAM chips.}
%\pagebreak
\section{Limitations}
\label{sec:limitations}
\micrev{We identify HiRA’s limitations under {{three}} categories.}

\micrev{First, we experimentally demonstrate that HiRA is supported by real DDR4 DRAM chips. However, we cannot verify the {\emph{exact}} {operation} of HiRA {(i.e., how HiRA is enabled)} in those chips for two reasons: 1) \emph{no} public documentation discloses or verifies HiRA in real DRAM chips and 2) we do \emph{not} have access to DRAM manufacturers’ proprietary circuit designs.}

\micrev{Second, all DRAM chips that exhibit successful HiRA operation are manufactured by {SK Hynix ({the second largest} DRAM manufacturer {that has \SI{27.4}{\percent} of {the} DRAM market share~\cite{hynixnambatwo}})}. We also conducted experiments using {40} DRAM chips from {{each of the} two} other manufacturers {(Samsung and Micron)} for which we observed \emph{no} successful HiRA operation. {We hypothesize that the DRAM chips from these other manufacturers \emph{ignore} the \gls{pre} or the second \gls{act} command of HiRA's command sequence when \gls{tras} and \gls{trp} timing parameters are greatly violated ({e.g.,} the DRAM chip acts as if it did not receive the \gls{pre} or the second \gls{act} commands).} Therefore, HiRA is currently limited to DRAM chips that {can successfully perform \gls{hira}} operation{s}. {We believe that other DRAM chips are fundamentally capable of \gls{hira} since \gls{hira} is consistent with fundamental operational principles of modern DRAM.} We hope that this work inspires future DRAM designs {that} explicitly support HiRA, given that HiRA 1) provides significant performance benefits and 2) is already {possible} in real DRAM chips, {even though DRAM chips are not even designed to support it}.}
\micrev{Third, performing periodic refresh using HiRA results in higher memory command bus utilization compared to using {conventional} {REF} commands. HiRA issues a row activation (ACT) {command} and a precharge (PRE) command to refresh a {\emph{{single}}} DRAM row, while \emph{multiple} DRAM rows are refreshed when a {single} REF command is issued. Even though HiRA overlaps the latency of row activation and precharge operations with the latency of other refresh or access operations, it still uses the command bus bandwidth to transmit {\gls{act} and \gls{pre}} commands to DRAM chips. {As the} number of ranks and banks per DRAM channel {increases}, HiRA’s command bus utilization can cause memory access requests to experience larger delays compared to using {REF commands} \sho{({as we evaluate in} \secref{sec:sensitivity})}. {However, \gls{hira} still provides \SI{12.1}{\percent} system performance benefit over a baseline memory controller that uses {REF commands} even in an 8-rank system {with high} command bus utilization.}}

{We {conclude} that none of these limitations fundamentally prevent a system designer from {using existing} DRAM chips {that can reliably perform HiRA operations} and {thus}, {benefit from} HiRA’s {refresh-refresh and refresh-access parallelization}.} 

\setstretch{0.94}
\section{Related Work}
\label{sec:relatedwork}

{To our knowledge, this is the first work {to} demonstrate {that} 1)~real off-the-shelf DRAM chips are capable of {refreshing a DRAM row concurrently with refreshing or activating another row within the same DRAM bank,}
by {carefully} violating {DRAM} timing parameters and 2) {doing so} is beneficial {to reduce the performance overhead of both {\emph{periodic}} refresh operations {(required for reliable DRAM operation)} and {\emph{preventive}} refresh operations {(required for RowHammer bit flip prevention)}.}}
We classify the related work into {six} {main} categories.

\noindent
{\textbf{Eliminating unnecessary refreshes~{(e.g., {\cite{liu2012raidr, lin2012secret, nair2013arch,baek2013refresh, qureshi2015avatar,patel2017reaper,venkatesan2006rapid,khan2017detecting,khan2016parbor,khan2014efficacy,ciji2009eskimo,song2011flikker,khan2016case, jung2015omitting,jafri2021refresh, katayama1999fault, wilkerson2010reducing, ghosh2007smart, song2000method, emma2008rethinking, mukhanov2019workload, hong2018ear, kraft2018improving, park2011power, wang2018content, kim2020charge}}).}}
{Various p}rior works eliminate unnecessary refresh operations by leveraging the heterogeneity in the retention time of DRAM cells to reduce the rate at which {some or all} DRAM rows are refreshed.
Our {work} differs from these {works} in {two} key aspects.
{First,} {most of} these works rely on identifying cells or rows with worst-case retention times, which is a difficult problem~\cite{mrozek2010analysis,mrozek2019multi, patel2017reaper, liu2012raidr, qureshi2015avatar, liu2013experimental}. {In contrast,} our work} uses a {relatively simple} {and well-understood} one-time {experiment} to {1)~identify \gls{hira}'s coverage and 2)~verify that \gls{hira} reliably works, presented in \secref{sec:experiments_coverage} and \secref{sec:experiments_chargerestoration}, respectively. Second, these works} focus on reducing the performance overhead of periodic refreshes, {but \emph{not} the increasingly worsening performance overhead of preventive refresh {operations} of RowHammer defenses~\rhdefworse{}. In contrast, our work tackles the performance overheads of} \emph{both} periodic and preventive refreshes.

\noindent
{\textbf{Circuit-level modifications to reduce the {performance impact} of refresh operation{s} {(e.g., \cite{hassan2019crow, luo2020clrdram, kim2000dynamic, kim2003block,yanagisawa1988semiconductor,ohsawa1998optimizing,nair2014refresh,orosa2021codic, choi2020reducing})}.}}
{{These} works develop DRAM-based techniques that 1)~reduce the latency of a refresh operation~\cite{hassan2019crow, luo2020clrdram}, {2)~implement a new refresh command that can be interrupted to quickly perform main memory accesses~\cite{nair2014refresh}}, and 3)~reduce the rate at which some or all DRAM rows are refreshed~\cite{kim2000dynamic,kim2003block,yanagisawa1988semiconductor,ohsawa1998optimizing}. {As opposed to \doduo{},} the{se techniques}} {1)} are \emph{not} compatible with {existing} DRAM chips as they require modifications {to} DRAM circuitry,
{and} 2) \emph{cannot} hide DRAM access latency in the presence of refresh operations.

\noindent
{\textbf{Memory {access} scheduling techniques to reduce the performance impact of refresh operations (e.g.,~\cite{mukundan2013understanding,stuecheli2010elastic, chang2014improving, pan2019colored, pan2019hiding, kotra2017hardware}).}}
{{Several} works {propose 
{issuing} $REF$ commands} during \emph{DRAM idle time} {(where no memory access requests {are} scheduled)} {to reduce the performance impact of refresh operations.} {{Most of these} works leverage} {the flexibility of delaying a $REF$ command for {multiple refresh intervals (e.g., for \SI{70.2}{\micro\second} in DDR4~\cite{jedec2017ddr4})}}. In contrast, \gls{hira} overlaps the latency of a refresh operation with other refreshes or memory accesses and thus can reduce the performance impact of refresh operations \emph{without} relying on {DRAM} idle time.} 

\noindent
{\textbf{Modifications to the DRAM architecture to leverage subarray-level parallelism {(e.g.,~\salprefs{})}.} Several works
partially overlap the latency of refresh operations or \memaccs{} via modifications to the DRAM architecture. {The \gls{hira} operation builds on the basic idea{s} of subarray-level parallelism introduced in~\cite{salp} and {refresh-access parallelization introduced in}~\cite{chang2014improving,zhang2014cream}}. {However,} unlike \doduo{}, these works require modifications to DRAM chip design, {and} thus they are \emph{not} compatible with {off-the-shelf} DRAM chips.} In contrast, {\doduo{}} uses {existing} {\gls{act} and \gls{pre}} commands{, {and we demonstrate that it works on real off-the-shelf DRAM chips}}.

\pagebreak
\noindent
{\textbf{RowHammer defense mechanisms {that use preventive refresh operations} {(e.g.,~\rhdefrefresh{})}.} {These works} propose} mechanisms that observe memory access patterns and speculatively schedule {preventive} refresh operations targeting potential victim rows of a RowHammer attack. {\doduo{} can be combined with all of these mechanisms to reduce the {performance} overhead{s of} preventive refresh {operations}.}

{\head{{Reducing the latency of {major} DRAM operations} {(e.g.,~\cite{hassan2016chargecache, das2018vrldram, wang2018caldram, zhang2016restore, shin2014nuat, lee2013tiered, lee2015adaptive, kim2018solar,chang2016understanding, chang2017understanding, lee2017design,chandrasekar2014exploiting,shin2015dram})}}} 
{{Many} {works}} {develop techniques that} {\emph{reduce}} the {latency of {major} DRAM operations (e.g., \gls{act}, \gls{pre}, \gls{rd}, \gls{wr}) by leveraging {1)}~temporal locality in workload access patterns~\cite{hassan2016chargecache,das2018vrldram,wang2018caldram,zhang2016restore,shin2014nuat}{,} 2)~the guardbands in manufacturer-recommended DRAM timing parameters~\cite{lee2013tiered,lee2015adaptive,kim2018solar,chang2016understanding,lee2017design, chang2017understanding}{, and 3)~variation in DRAM {latency} due to temperature dependenc{e}~\cite{chandrasekar2014exploiting,lee2015adaptive,kim2018solar}.}}
{{These latency reduction techniques} can improve system performance by alleviating the performance impact of refresh operations (e.g., a refresh can be performed {faster} with a reduced charge restoration latency).}
{These techniques can be combined with \doduo{} to further alleviate the performance impact of refresh operations, as \doduo{} \emph{overlaps} 
the latency of refreshing a DRAM row 
with the latency of refreshing or {activating} another row {in the same bank}.}

\section{Conclusion}
\label{sec:conclusion}

{We introduce} \gls{hira}, {a new {DRAM} operation} {that can} {reliably parallelize a DRAM row's refresh operation with refresh or activation of another row within the same bank.}
\gls{hira} achieves this by {activating} two electrically-isolated rows {in quick succession,} allowing them to be refreshed{/{activated}} without disturbing each other.
{{W}e show that \gls{hira} 1)~works reliably in 56 real off-the-shelf DRAM chips}{, using already{-}available (i.e., standard) \gls{act} and \gls{pre} DRAM commands{,} by violating timing constraints and 2)~reduces} the {overall} latency {of refreshing two rows} by \SI{51.4}{\percent}.
To leverage {the parallelism} \gls{hira} {provides,}
we {design} \glsfirst{hirasched}.  
\gls{hirasched} modifies the memory request scheduler to perform \gls{hira} operations when a periodic or {RowHammer-preventive} refresh can be {performed} concurrently with another {refresh {or row activation}} {to} the same bank. {{Our system-level evaluations} show that} {\gls{hirasched}} {increases system performance by \SI{12.6}{\percent} and $3.73\times$ {{as it reduces}}}
the performance {degradation {due to}} periodic and {preventive} refreshes{,}
respectively.
{We conclude that \gls{hira} {{1)~}already works in off-the-shelf DRAM chips and can be used to} significantly reduce the {performance degradation caused by} both periodic and preventive refreshes {and 2)~provides higher performance benefits in higher{-}capacity DRAM chips.}}
{We hope that our findings will inspire DRAM {manufacturers and} {standards bodies} to {explicitly} {and properly} support \gls{hira} in future DRAM {chips and standards}.}

\section*{Acknowledgments}
We thank our shepherd and the reviewers of ASPLOS'22, ISCA'22, and MICRO'22 for valuable feedback. We thank the SAFARI Research Group members for {useful} feedback and the stimulating intellectual environment they provide. We acknowledge the generous gifts provided by our industrial partners{, including} Google, Huawei, Intel, Microsoft, and VMware{, and support from {the} Microsoft Swiss Joint Research Center}. 
\pagebreak
% \clearpage
\balance
\bibliographystyle{IEEEtranS}
\bibliography{main}

% \ifarb
\onecolumn
\appendix

\setcounter{table}{3}
\section{Tested DRAM Chips}
\label{sec:appendix_chips}
\newcommand*{\myalign}[2]{\multicolumn{1}{#1}{#2}}

\agyarb{Table~\ref{tab:detailed_info} shows the characteristics of the DDR4 DRAM modules we test and analyze. {We provide {the} access frequency (Freq.), manufacturing date (Date Code), {chip} capacity (Chip Cap.), die revision {(Die Rev.)}, and chip organization {(Chip Org.)} of tested DRAM modules. We report the manufacturing date of these modules in the form of $week-year$. For each DRAM module, Table~\ref{tab:detailed_info} shows two \gls{hira} characteristics in terms of minimum (Min.), average (Avg.) and maximum (Max.) values across all tested rows: 
1)~HiRA Coverage: the fraction of DRAM rows
within a bank which HiRA can reliably activate concurrently
with refreshing a given row (\secref{sec:experiments_coverage})
and 2)~Norm. $N_{RH}$: the increase in the RowHammer threshold when \gls{hira}'s second row activation is used for refreshing the victim row (\secref{sec:experiments_chargerestoration}).}}
\setlength{\tabcolsep}{5.5pt}
\begin{table*}[ht]
\footnotesize
\centering
\caption{Characteristics of the tested DDR4 DRAM modules.}
\label{tab:detailed_info}
\begin{tabular}{l|l|l||ccccc|ccc|ccc}
\textbf{Module} & \textbf{Module} & \textbf{Module Identifier} & \textbf{Freq} & \textbf{Date} & \textbf{Chip} & \textbf{Die} & \textbf{Chip} & \multicolumn{3}{c|}{\textbf{HiRA Coverage}} & \multicolumn{3}{c}{\textbf{Norm. $N_{RH}$}} \\ 
\textbf{Label}&\textbf{Vendor}&\textbf{Chip Identifier}&\textbf{(MT/s)}&\textbf{Code}&\textbf{Cap.}&\textbf{Rev.}&\textbf{Org.}&\textbf{Min.}&\textbf{Avg.}&\textbf{Max.}&\textbf{Min.}&\textbf{Avg.}&\textbf{Max.}\\
\hline
\hline
A0 & \multirow{2}{*}{G.SKILL} & \multirow{2}{*}{\begin{tabular}[l]{@{}l@{}}DWCW (Partial Marking)$^*$\\F4-2400C17S-8GNT~\cite{datasheetF42400C17S8GNT}\end{tabular}} & \multirow{2}{*}{2400} & \multirow{2}{*}{42-20} & \multirow{2}{*}{4Gb} & \multirow{2}{*}{B} & \multirow{2}{*}{x8} & \SI{24.8}{\percent} & \SI{25.0}{\percent} & \SI{25.5}{\percent} & 1.75&1.90&2.52\\
A1 & & & & & & & & \SI{24.9}{\percent} & \SI{26.6}{\percent} & \SI{28.3}{\percent} & 1.72 & 1.94 & 2.55 \\ \hline
B0 & \multirow{2}{*}{Kingston} & \multirow{2}{*}{\begin{tabular}[l]{@{}l@{}}H5AN8G8NDJR-XNC\\KSM32RD8/16HDR~\cite{datasheetksm32rd8}\end{tabular}} & \multirow{2}{*}{2400} & \multirow{2}{*}{48-20} & \multirow{2}{*}{4Gb} & \multirow{2}{*}{D} & \multirow{2}{*}{x8} & \SI{25.1}{\percent} & \SI{32.6}{\percent} & \SI{36.8}{\percent} & 1.71&1.89&2.34\\
B1 & & & & & & & & \SI{25.0}{\percent} & \SI{31.6}{\percent} & \SI{34.9}{\percent} & 1.74&1.91&2.51\\ \hline
C0 & \multirow{3}{*}{SK Hynix} & \multirow{3}{*}{\begin{tabular}[l]{@{}l@{}}H5ANAG8NAJR-XN\\HMAA4GU6AJR8N-XN~\cite{datasheetHMAA4GU6AJR8N}\end{tabular}} & \multirow{3}{*}{2400} & \multirow{3}{*}{51-20} & \multirow{3}{*}{4Gb} & \multirow{3}{*}{F} & \multirow{3}{*}{x8} & \SI{25.3}{\percent} & \SI{35.3}{\percent} & \SI{39.5}{\percent} & 1.47 & 1.89 & 2.23 \\ 
C1 & & & & & & & & \SI{29.2}{\percent} & \SI{38.4}{\percent} & \SI{49.9}{\percent} & 1.09 & 1.88 & 2.27\\ 
C2 & & & & & & & & \SI{26.5}{\percent} & \SI{36.1}{\percent} & \SI{42.3}{\percent} & 1.49 & 1.96 & 2.58\\ \hline
\hline
\end{tabular}
\begin{flushleft}
$^*$ The chip identifier is partially removed on these modules. We infer the chip manufacturer and die revision based on the remaining part of the chip identifier.

\end{flushleft}
\end{table*}

\end{document}